\documentclass[twocolumn,showpacs,preprintnumbers,amsmath,amssymb]{revtex4}

\DeclareMathAlphabet{\mathpzc}{OT1}{pzc}{m}{it}

\def\be{\begin{equation}}
\def\ee{\end{equation}}

\def\ba{{\bf a}}
\def\bk{{\bf k}}
\def\br{{\bf r}}
\def\bu{{\bf u}}

\def\bx{{\bf x}}
\def\by{{\bf y}}

\def\bA{{\bf A}}
\def\bB{{\bf B}}

\def\bG{{\bf G}}

\def\bW{{\bf W}}
\def\wh{\widetilde}
\def\lb{\label}

\def\bar{\overline}

\def\brho{\hbox{\boldmath $\rho$}}
\def\bdelta{\hbox{\boldmath $\delta$}}
\def\bdot{\hbox{\boldmath $\cdot$}}
\def\btimes{\hbox{\boldmath $\times$}}

\def\bzed{\hbox{\boldmath $0$}}
\def\grad{\hbox{\boldmath $\nabla$}}
\def\bell{\hbox{\boldmath $\ell$}}

\begin{document}

\preprint{APS/123-QED}

\title{Small-Scale Kinematic Dynamo and Non-Dynamo in Inertial-Range
Turbulence}

\author{Gregory L. Eyink}
\email{eyink@ams.jhu.edu}
\affiliation{%
Department of Applied Mathematics \& Statistics\\
The Johns Hopkins University, USA}%

\author{Ant\^{o}nio F. Neto}
\email{antfrannet@gmail.com}
\affiliation{Campus Alto Paraopeba\\
Universidade Federal de S\~{a}o Jo\~{a}o del-Rei, Brazil}%

\date{\today}

\begin{abstract}
We investigate the Lagrangian mechanism of the kinematic ``fluctuation''
magnetic
dynamo in turbulent plasma flow at small magnetic Prandtl numbers. The combined
effect of turbulent advection and plasma resistivity is to carry infinitely
many
field lines to each space point, with the resultant magnetic field at that
point given
by the average over all the individual line vectors. As a consequence of the
roughness
of  the advecting velocity, this remains true even in the limit of zero
resistivity. We show
that the presence of dynamo effect requires sufficient angular correlation of
the passive
line-vectors that arrive simultaneously at the {\it same} space point. We
demonstrate
this in detail for the Kazantsev-Kraichnan model of kinematic dynamo with a
Gaussian
advecting velocity that is spatially rough and white-noise in time. In the
regime where
dynamo action fails, we also obtain the precise rate of decay of the magnetic
energy.
These exact results for the model are obtained by a generalization of the
``slow-mode
expansion'' of Bernard, Gaw\c{e}dzki and Kupiainen to non-Hermitian evolution.
Much
of our analysis applies also to magnetohydrodynamic turbulence.
\end{abstract}


\maketitle

\section{\label{sec:level1}Introduction}

Turbulent magnetic dynamo effect is of great importance in astrophysics and
geophysics \cite{BrandenburgSubramanian05}. Many questions remain, however,
about the basic mechanism of dynamo action, even for the kinematic stage when
the seed magnetic field is weak and does not react back on the advecting
velocity field.
Stretching of field lines by a chaotic flow is, of course, the ultimate source
of growth
of magnetic field strength.  Plasma resistivity $\eta$ in turn acts to damp the
magnetic field. However, the dynamo cannot be understood as a simple
competition
between growth from stretching and  dissipation from resistivity. For example,
resistivity
plays also a positive role in dynamo effect through the reconnection of
complex,
small-scale field-line structure \cite{Moffatt78}.

In addition, random advection may not lead to field growth in the limit of
vanishing resistivity.
Consider, for instance, the kinematic dynamo model of Kazantsev
\cite{Kazantsev68}
and Kraichnan \cite{KraichnanNagarajan67,Kraichnan68} with a Gaussian random
velocity
that is delta-correlated in time. In this model there is a dramatic dependence
of dynamo
effect on the spatial rugosity of the velocity, as measured by the scaling
exponent $0<\xi<2$
of the spatial 2-point velocity correlation \cite{Kazantsev68}. There exists a
certain critical
value $\xi_*$ such that for $\xi<\xi_*,$ kinematic dynamo effect exists only
above a threshold
value $Pr_c$ of the magnetic Prandtl number $Pr=\nu/\kappa$ \cite{Vincenzi02,
ArponenHorvai07}. Here $\kappa=\eta c/4\pi$ is magnetic diffusivity while $\nu$
is an
effective viscosity associated to a ``dissipation length'' $\ell_\nu$ of the
velocity, above
which scaling holds with exponent $\xi$ and  below which the synthetic field
becomes
perfectly smooth.  In this regime of extreme roughness of the advecting
velocity field,
there is no kinematic dynamo even as $\kappa,\nu\rightarrow 0,$ if $Pr<Pr_c.$

On the contrary, in the Kazantsev-Kraichnan (KK) model for smoother velocity
fields with
$\xi>\xi_*$ there is a critical value $Re_{m,c}$ of the magnetic Reynolds
number
$Re_m= u_{rms}L/\kappa,$ where $u_{rms}$ is the root-mean-square velocity
and $L$ is the integral length-scale of the fluctuating velocity. See
\cite{Vincenzi02,
ArponenHorvai07}, also \cite{BoldyrevCattaneo04}. In this smooth regime,
small-scale kinematic dynamo leads to exponential growth of the rms magnetic
field, even
for $Pr\rightarrow 0$ as long as $Re_m>Re_{m,c}.$ The most natural
correspondence of the KK model to the kinematic dynamo problem in real fluid
turbulence is for the value $\xi=4/3,$ which is greater than the critical value
$\xi_*=1$
in three space dimensions. This correspondence would suggest that there is a
critical
magnetic Reynolds number for onset of kinematic dynamo in actual fluid
turbulence,
but no lower bound on the magnetic Prandtl number. On the other hand, numerical
studies of Schekochihin et al. \cite{Schekochihinetal04,Schekochihinetal05}
suggested
that hydrodynamic turbulence is instead like the rough regime of the KK model
and
that the critical Prandtl number $Pr_c$ for small-scale dynamo action tends to
a finite,
positive value as $Re_m\rightarrow\infty.$  Their latest investigations now
support the
opposite conclusion, that the critical magnetic Reynolds number $Re_{m,c}$
tends to
a finite, positive value as $Pr\rightarrow
0$\cite{Iskakovetal07,Schekochihinetal07}.
Considerable debate still continues, however, about the precise nature and
universality
of the observed small-scale dynamo.

To resolve such subtle issues a better physical understanding is required of
the mechanism
of the turbulent kinematic dynamo. In our opinion, important ideas have been
contributed
recently by Celani et. al. \cite{Celanietal06}. They pointed out that the
existence of
dynamo effect in the KK model for space dimension $d=3$ should be closely
related
to the angular correlation properties of material line-vectors. They considered
 the
covariance at time $t$ of two infinitesimal line-vectors that are advected
starting a distance
$r$ apart at time $0.$ Celani et al. argued that this correlation vanishes as
$r$
decreases through the inertial scaling range or as  $t\rightarrow\infty,$ going
to zero
as a power $(r/t^{1/\gamma})^{\bar{\zeta}}.$ Here $\gamma=2-\xi$ and
$\bar{\zeta}
=\bar{\zeta}(\xi)$ is the scaling exponent of a ``homogeneous zero-mode'' for
the linear
operator ${\cal M}_2^*$ that evolves the pair correlations of line-elements
forward in time.
Celani et. al. \cite{Celanietal06} further claimed that the transition between
dynamo regimes
in the KK model for $d=3$ corresponds exactly to the value $\xi_*=1$ where
$\bar{\zeta}(\xi_*)=0$ \cite{Celanietal06}.

In this paper we shall further investigate these questions. In the first place,
we shall
show that the claims of Celani et. al. \cite{Celanietal06} are not quite
correct. It will be
shown here that the specific correlation function proposed by those authors
does
{\it not} discriminate between dynamo and non-dynamo regimes. The scaling law
which
they proposed is valid, but holds over the entire range $0<\xi<2$ with a
different zero-mode
and different scaling exponent than they had claimed. We shall show that a
quite different
correlation function of material line-elements is necessary to serve as an
``order parameter''
for kinematic dynamo. The crucial difference is that the quantity introduced
here measures
the angular correlation of material line-vectors that are advected to the {\it
same} space
point at time $t$. But still, why should there be any connection of roughness
exponent $\xi$
with dynamo action? Individual field lines ought to be stretched and their
field strengths
increased for all values of $\xi.$ We shall provide in this work a plausible
physical
explanation. Although individual lines may stretch due to chaotic advection,
infinitely-many
magnetic field lines will arrive at each point of the fluid due to diffusion by
resistivity
and the final magnetic field will be the average value that results from
reconnection
and  ``gluing'' of field lines by resistivity. We shall show that too little
angular correlation
leads to large cancellations in this resistive averaging, with the net magnetic
field
suffering decay despite the growth of individual field lines.

We devote the remainder of our paper to a detailed study of the ``failed dynamo
regime'' in the KK model for $\xi<\xi_*$ and $Pr<Pr_c.$ Part of our motivation
is
the speculation of \cite{Schekochihinetal04,Schekochihinetal05} that
hydrodynamic
turbulence at high magnetic Reynolds number but low $Pr$ resembles this
parameter
range of the KK model. A better understanding of this regime may be useful to
{\it rule out}
its validity for hydrodynamic turbulence, based on astrophysical observations.
We shall
see, for example, that it implies a very rapid rate of decay of an initial seed
magnetic field.
Indeed, we show that in the KK model for rougher velocities the decay of the
magnetic
field is not resistively limited, with dissipation rate non-vanishing even in
the zero-resistivity
limit $\kappa\rightarrow 0,$ as long as $Pr<Pr_c.$ There is a strong analogy
with the
anomalous decay of a turbulence-advected passive scalar, for which scalar
dissipation
is non-vanishing even in the limit of zero scalar
diffusivity\cite{EyinkXin00,Chavesetal01}.
The decay rate is instead determined by large-scale statistical conservation
laws,
associated to ``slow modes'' of the scalar evolution operator. We show here
that the
decay of the magnetic field in the rough regime of the KK model is determined
in the
limit  $\kappa\rightarrow 0,\,\,Pr<Pr_c$ by the ``slow modes'' of the linear
evolution
operator ${\cal M}_2^*$ for pairs of infinitesimal line-elements.  We shall
establish
these results by a formal extension of the slow-mode expansion of Bernard et
al.
\cite{Bernardetal97} to the case of non-Hermitian evolution operators, which is
presented in the Appendix.  We shall furthermore determine all self-similar
decay
solutions of the magnetic field in the non-dynamo regime of the KK model,
following
\cite{EyinkXin00} for the passive scalar. Unlike the scalar case, however,
determining
the decay law of the magnetic energy requires an additional step of matching
these
self-similar solutions to explicit resistive-range solutions. We shall use
these results to discuss
the physical mechanism of kinematic dynamo, and, in particular, to relate our
dynamo
``order parameter''  to the process of ``induction'' by a spatially uniform
initial magnetic
field.  As we shall see,  considerable insight can be obtained into the inner
workings
of the small-scale dynamo by considering also the situations where it fails.

%


\section{\label{sec:level2}The Kraichnan-Kazantsev Dynamo
and Correlations of Line-Elements}

\subsection{The Kinematic Dynamo}\lb{kin-dyn}

The evolution of the passive magnetic field $\bB(\bx,t)$ is governed by
the induction equation
\begin{equation}\label{MHD}
\partial_t\bB+(\bu\bdot\grad)\bB
-(\bB\bdot\grad)\bu=\kappa\triangle\bB,
\end{equation}
where $\bu\equiv \bu(\bx,t)$ is the advecting velocity field and $\kappa$ is
the
magnetic diffusivity. The magnetic field is taken to  be solenoidal, assuming
there are no magnetic monopoles:
\begin{equation}\label{solenoid}
\grad\bdot\bB=0.
\end{equation}
Notice that this condition is preserved by the evolution equation (\ref{MHD})
if it
is imposed at the initial time $t_0=0.$ We have also assumed above that the
advecting fluid is incompressible so that
\begin{equation}\label{incompress}
\grad\bdot\bu=0. \\
\end{equation}
For simplicity, we shall only consider this case hereafter.

For an incompressible fluid, one can represent the solution of the
induction equation by a stochastic Lagrangian representation of the
following form:
\begin{equation}
\mathbf{B}(\mathbf{x},t)=\mathbb{E} \left[
\mathbf{B}_0(\mathbf{a})\bdot\grad_\mathbf{a} \tilde
{\mathbf{x}}(\mathbf{a},t)\,\vrule_{\mathbf{a}=\tilde{\mathbf{a}}
(\mathbf{x},t)}\right].
\end{equation}
See \cite{Celanietal06,Eyink09}.
Here $\tilde {\ba}(\bx,t)$ are the ``back-to-label maps''
for stochastic forward flows $\tilde {\bx}(\ba,t)$ solving the SDE
\begin{equation}\label{Eqx(a,t)}
d\tilde{\bx}(\ba,t)=\bu(\tilde{\bx}(\ba,t),t)dt+\sqrt{2\kappa}\, d\bW(t).
\end{equation} $\mathbb{E}$ denotes average over the realizations
of the Brownian motion $\bW(t)$ in Eq. (\ref{Eqx(a,t)}). For the
$i$th component of the magnetic field we can write
\begin{equation}
\begin{array}{lll}
B^i(\mathbf{x},t)&=&\mathbb{E}\left[B_{0}^{k}(\mathbf{a})
\displaystyle\frac{\partial \tilde{x}^i}{\partial a^k}
(\mathbf{a},t)|_{\mathbf{a}=\tilde{\mathbf{a}}(\mathbf{x},t)}\right]
\vspace{2mm}\\
&=&\int d^da\, B_{0}^{k}(\mathbf{a})
\hat{F}_k^i\left(\mathbf{a},0|\mathbf{x},t;\mathbf{u}\right)
\end{array}
\end{equation} where we have defined
\begin{equation}
\hat{F}_k^i\left(\mathbf{a},0|\mathbf{x},t;\mathbf{u}\right)
\equiv \mathbb{E}\left[\displaystyle\frac{\partial
\tilde{x}^i}{\partial
a^k}(\mathbf{a},t)\delta^d\left(\mathbf{a}-\tilde{\mathbf{a}}
(\mathbf{x},t)\right)\right]
\end{equation}

These results may be used to represent the equal-time, 2-point correlation
of the magnetic field, averaged over the stochastic velocity
$\mathbf{u}$ and the random initial magnetic field $\mathbf{B}_0:$
$$\begin{array}{lll}
\left<B^i(\mathbf{x},t)B^j(\mathbf{x}',t)\right>&=&\int d^d a\int
d^da' \left<B_{0}^{k}(\mathbf{a})B_{0}^{\ell}(\mathbf{a}')\right>
\vspace{2mm}\\
&& \hspace{4mm}\bar{F}^{ij}_{k\ell}
\left(\mathbf{a},\mathbf{a}',0|\mathbf{x},\mathbf{x}',t\right).
\end{array}$$
We have assumed that $\mathbf{u}$ and $\mathbf{B}_0$ are
statistically independent and we also have defined
\begin{equation}
\bar{F}^{ij}_{k\ell}
\left(\mathbf{a},\mathbf{a}',0|\mathbf{x},\mathbf{x}',t\right)=
\left<\hat{F}^{i}_{k}(\mathbf{a},0|\mathbf{x},t)
\hat{F}^{j}_{\ell} (\mathbf{a}',0|\mathbf{x}',t)\right>.
\end{equation}
For statistically homogeneous velocity and initial conditions, with
$\mathcal{C}^{ij}(\mathbf{r},t)\equiv
\left<B^{i}(\mathbf{x},t)B^{j}(\mathbf{x}',t)\right>$ for
$\mathbf{r}=\mathbf{x}-\mathbf{x}'$, we obtain
\begin{equation}
\mathcal{C}^{ij}(\mathbf{r},t) =\int d^d\rho \,\,\mathcal{C}^{k\ell}(\brho,0)
\bar{F}^{ij}_{k\ell}\left(\brho,0|\mathbf{r},t\right)
\lb{C-prop} \end{equation}
with
\begin{eqnarray}
\bar{F}^{ij}_{k\ell}\left(\brho,0|\mathbf{r},t\right)
&=&\mathbb{E}\mathbb{E}'\left[ \left< \displaystyle\frac{\partial
\tilde{x}^i}{\partial
a^k}(\mathbf{a}+\brho,t)\displaystyle\frac{\partial
\tilde{x}^{\prime j}}{\partial a^{\ell}}(\mathbf{a},t)\Bigg|_{\mathbf{a}
=\mathbf{\tilde{a}}'(\mathbf{x},t)}\right.\right.\vspace{2mm}\cr
&&\left.\times\delta^d\left(
\tilde{\ba}(\mathbf{x}+\mathbf{r},t)
-\tilde{\ba}'(\mathbf{x},t)-\brho\right)\Bigg\rangle\right].
\lb{Fbar-def}\end{eqnarray}
Here the prime $\prime$ denotes a second Brownian motion $\bW'(t)$
statistically independent of $\bW(t).$ Equation (\ref{C-prop}) was introduced
by Celani et al. \cite{Celanietal06} and heavily exploited in their analysis
of the magnetic correlation.

Another closely related propagator was introduced by \cite{Celanietal06}
related to infinitesimal material line-elements, which evolve according
to the Lagrangian equation of motion:
$$ D_t \delta\bell = (\delta\bell\bdot\grad)\bu. $$
Note that the positions of the line-elements are assumed to move stochastically
according to (\ref{Eqx(a,t)}), so these are not quite ``material lines'' in the
traditional sense when $\kappa>0.$ The exact solution of the above equation
for $t>0$ is
$$ \delta\bell(t)=\delta\bell(0)\bdot\grad_\mathbf{a} \tilde
{\mathbf{x}}(\mathbf{a},t), $$
with $\tilde{\bx}(\mathbf{a},t)$ solving (\ref{Eqx(a,t)}).
Taking initial line-elements $\delta\ell^i_k(0)=\delta^i_k,\,
\delta\ell^{\prime j}_\ell(0)=\delta^j_\ell$ starting at positions $\ba,\ba'$
displaced  by $\br=\ba'-\ba,$ one may follow \cite{Celanietal06} to define for
statistically homogeneous turbulence
\begin{eqnarray}
F^{ij}_{k\ell}(\brho,t|\br,0) &=&
\langle \delta\ell^i_k(t)\delta\ell^{\prime j}_\ell(t)\delta^d(\tilde{\bx}(t)-
  \tilde{\bx}'(t)-\brho)\rangle\cr
&=&\mathbb{E}\mathbb{E}'\left[
 \left<\displaystyle\frac{\partial \tilde{x}^i}{\partial
a^k}(\mathbf{a}+\br,t)\displaystyle\frac{\partial
\tilde{x'}^j}{\partial a^{\ell}}(\mathbf{a},t)\right.\right.\vspace{2mm}\cr
&&\left.\times\delta^d\left( \tilde{\bx}(\mathbf{a}+\br,t)
-\tilde{\bx}'(\ba,t)-\brho\right)\Bigg\rangle\right]. \vspace{2mm}
\lb{F-def} \end{eqnarray}
If we make the change of variables $\ba \mapsto
\bx$ in the argument of the delta function of eq.(\ref{Fbar-def}), then the
jacobian
of this transformation of variables is 1 due to incompressibility. Therefore,
one
finds by comparison with (\ref{F-def}) that
$$\begin{array}{lll}
\bar{F}^{ij}_{k\ell}\left(\brho,0|\br,t\right)
&=&\mathbb{E}\mathbb{E}'\left[
 \left<\displaystyle\frac{\partial \tilde{x}^i}{\partial
a^k}(\mathbf{a}+\brho,t)\displaystyle\frac{\partial
\tilde{x'}^j}{\partial a^{\ell}}(\mathbf{a},t)\right.\right.\vspace{2mm}\\
&&\left.\times\delta^d\left( \tilde{\bx}(\mathbf{a}+\brho,t)
-\tilde{\bx}'(\ba,t)-\br\right)\Bigg\rangle\right]\vspace{2mm}\\
&=&F^{ij}_{k\ell}(\br,t|\brho,0),
\end{array}$$
equating the two propagators under interchange of arguments.

In our work below an important role will also be played by the covariant
vector given by the gradient $\bG=\grad\theta$ of a passive scalar
$\theta$. The  gradient satisfies the equation
\begin{equation}\label{Geq}
\partial_t\bG+(\bu\bdot\grad)\bG
+(\grad\bu)\bG=\kappa\triangle\bG.
\end{equation}
which is dual to the equation (\ref{MHD}) for the contravariant vector $\bB$
\cite{Saffman63}. The above equation preserves the condition $\bG=\grad\theta$
if this is imposed at time $t_0=0.$ A stochastic Lagrangian representation also
exists for the solution of this equation. Solved forward in time with
$\kappa>0$
this representation involves the matrix $\grad_\bx\tilde{\ba}(\bx,t).$ However,
taking $\kappa\rightarrow-\kappa$ in (\ref{Geq}) and solving backward
from time $t>0$ to time $0$ yields the representation
for the $i$ component of the gradient field:
\begin{eqnarray*}
G_i(\ba,0)&=&\mathbb{E}\left[
\displaystyle\frac{\partial \tilde{x}^k}{\partial a^i}(\mathbf{a},t)
G_k(\tilde{\bx}(\ba,t),t) \right] \cr
  &=& \int d^dx\, G_k(\mathbf{x},t)
\hat{F}_i^k\left(\mathbf{a},0|\mathbf{x},t;\mathbf{u}\right).
\end{eqnarray*}
For statistically homogeneous velocity and initial conditions we
introduce the 2-point correlation of the gradient field, $\mathcal{G}_{ij}
(\brho,t)\equiv \left<G_i(\ba,0)G_j(\ba',0)\right>$ with $\brho=\ba-\ba'.$
By the same arguments as previously
\begin{eqnarray}
\mathcal{G}_{ij}(\brho,0) &= & \int d^dr \,\mathcal{G}_{k\ell}(\br,t)
           \bar{F}^{k\ell}_{ij}\left(\brho,0|\br,t\right), \cr
           &= & \int d^dr \,\mathcal{G}_{k\ell}(\br,t)
           F^{k\ell}_{ij}\left(\br,t|\brho,0\right),
\lb{G-prop} \end{eqnarray}
for positive times $t>0.$


We shall generally avoid using the geometric language of differential forms
and Lie derivatives in this paper, but a few brief remarks may be useful.
For those unfamiliar with this formalism, a good introductory reference is
\cite{Larsson03}. The magnetic field $\bB$ discussed above is a 1-form,
which is more properly represented by its Hodge dual $\bB^*,$ a $(d-1)$-form.
The equation (\ref{MHD}) for $\kappa=0$ is equivalent to $\partial_t\bB^*+
\mathcal{L}\bB^*=0,$ where $\mathcal{L}$ is the Lie derivative. The Lie
derivative
theorem thus implies that (\ref{MHD}) for $\kappa=0$ satisfies an analogue of
the
Alfv\'{e}n theorem, with conserved flux of $\bB$ through $(d-1)$-dimensional
material hypersurfaces. On the other hand, the field $\bG$ is a proper
1-form and equation (\ref{Geq}) for $\kappa=0$ is equivalent to $\partial_t\bG+
\mathcal{L}\bG=0.$ The Lie derivative theorem thus implies that integrals of
$\bG$
along material  lines are conserved for $\kappa=0.$ Either of these equations
could be regarded as a valid generalization of the kinematic dynamo problem
to general space dimension $d.$ The non-gradient solutions $\bG$ of (\ref{Geq})
are generalizations of the 3-dimensional vector potential and the ``magnetic
flux'' is represented by their line-integrals around closed material loops.
For $d=3$ the non-gradient solutions of (\ref{Geq}) are in one-to-one
correspondence (up to gauge transformations) with the solenoidal solutions
of (\ref{MHD}), by the familiar relation $\bB=\grad\btimes\bG.$

We note finally that all of the results in this section hold for any random
velocity
field that is divergence-free and statistically homogeneous. Thus, they apply
not only to the Kazantsev-Kraichnan model discussed in the following sections,
but also to the kinematic dynamo problem for hydrodynamic turbulence
governed by the incompressible Navier-Stokes equation.

\subsection{White-Noise Velocity Ensemble}\lb{white-noise}

In the Kazantsev-Kraichnan (KK) model \cite{Kazantsev68,KraichnanNagarajan67,
Kraichnan68} the advecting velocity $\bu(\bx,t)$ is taken to be a Gaussian
random field
with zero mean and second-order correlation delta function in time, given
explicitly by
\begin{equation}
\left<u^i(\bx,t)u^j(\bx',t')\right>=
\left[D_0\delta^{ij}-S^{ij}(\br)\right]\delta(t-t') \lb{u-cov}
\end{equation} with $\br=\bx'-\bx$. Under the incompressibility
constraint $\partial_{i}S^{ij}(\br)=0$ and supposing $S^{ij}(\br)$ scales
as $\sim r^{\xi}$ for $\ell_\nu\ll r\ll L,$ one deduces for that range that
\begin{equation}
S^{ij}(\br)=D_1 r^{\xi}
\left[\left(\xi+d-1\right)\delta^{ij}-\xi\hat{r}^{i}\hat{r}^{j}\right]
\lb{vel-corr}
\end{equation} where $\hat{r}^{i}=r^{i}/r$. Define ``viscosity''
$\nu=D_1\ell_\nu^\xi.$
Below we shall consider especially the limit $\nu,\kappa\rightarrow 0$ with
$\nu<
Pr_c \kappa.$ This is the non-dynamo regime in the limit of infinite
kinetic and magnetic Reynolds numbers. One of our main objectives
is to understand better the geometric and statistical properties of this
regime which lead to the failure of small-scale dynamo action.

In addition to the properties of statistical homogeneity, stationarity and
incompressibility, the white-noise velocity ensemble is time-reflection
symmetric.
This implies
$$ \bar{F}^{ij}_{k\ell}\left(\br,-t|\brho,0\right) =
\bar{F}^{ij}_{k\ell}\left(\br,t|\brho,0\right) $$
and the similar property for $F^{ij}_{k\ell}\left(\brho,t|\br,0\right).$
Combined with the
other symmetries, this imples that
$$\begin{array}{lll}
F^{ij}_{k\ell}\left(\brho,t|\br,0\right)
&=&\bar{F}^{ij}_{k\ell}\left(\br,0|\brho,t\right)
\vspace{2mm}\\
&=&\bar{F}^{ij}_{k\ell}\left(\br,-t|\brho,0\right)
\vspace{2mm}\\
&=&\bar{F}^{ij}_{k\ell}\left(\br,t|\brho,0\right).
\end{array}$$
The first line follows from incompressibility, the second line is due to
time-translation
invariance and the last equality follows from time-reflection symmetry. Thus
the two
propagators are adjoints in the KK model.

Time-reflection symmetry has also an important implication for the evolution
of the gradient field correlation. Note that time-translation invariance
implies
that equation (\ref{G-prop}) can be written for $t>0$ as
$$ \mathcal{G}_{ij}(\brho,-t) = \int d^dr \,\mathcal{G}_{k\ell}(\br,0)
           F^{k\ell}_{ij}\left(\br,0|\brho,-t\right).$$
Then time-reflection symmetry implies further that
\be \mathcal{G}_{ij}(\brho,t) = \int d^dr \,\mathcal{G}_{k\ell}(\br,0)
           F^{k\ell}_{ij}\left(\br,0|\brho,t\right) \lb{G-prop2} \ee
for $t>0.$ Compare with equation (\ref{C-prop}) for the magnetic correlation.
We see that the $F$-propagator in the KK model evolves forward in time the
gradient correlation.

The most important property of the white-noise model is its Markovian
character, which implies that time-evolution of correlations is governed
by  2nd-order differential (diffusion) equations. E.g. the $n$-th order
equal time correlation function $\mathcal{C}_{n}^{i_1i_2\ldots i_n}\equiv
\left<\prod_{a=1}^{n}B^{i_a}(\bx_a,t)\right>$ satisfies an equation
of the form $\partial_t\mathcal{C}_n=\mathcal{M}_n\mathcal{C}_n.$
Expressions for the general $n$-body diffusion operators $\mathcal{M}_n$
can be found in \cite{RogachevskiiKleeorin97}, which can be obtained using
It\^o formula as in Ref. \cite{ArponenHorvai07} or, equivalently, by Gaussian
integration by parts. In the limit $\nu,\kappa\rightarrow 0$
all of these operators for general $n$ become degenerate (singular)
and homogeneous of degree $-\gamma$ with $\gamma=2-\xi.$  Below we
shall mainly consider $n=2$ and thus write simply $\cal M$ for ${\cal M}_{2}.$
However, many of our considerations carry over to general $n,$ as will be
noted explicitly below. Following the notations of \cite{Celanietal06}, we
write
for $n=2$:
\be \partial_t\mathcal{C}^{ij}(\mathbf{r},t)
      = \left[\mathcal{M}(\br)\right]_{pq}^{ij}\mathcal{C}^{pq}(\mathbf{r},t)
\lb{C-eq} \ee
with
\be \left[\mathcal{M}\right]^{ij}_{pq}=\delta^i_p\delta^j_q
S^{\alpha\beta}\partial_\alpha\partial_\beta
-\delta^i_p \partial_qS^{\alpha j}\partial_\alpha
-\delta^j_q \partial_pS^{i\beta}\partial_\beta
+\partial_p\partial_q S^{ij}. \lb{M-def} \ee
The notation $\mathcal{M}(\br)$ indicates that $\partial_i=\partial/\partial
r^i.$
Note that equation (\ref{C-eq}) has an invariant subspace satisfying
$\partial_i\mathcal{C}^{ij}=\partial_j\mathcal{C}^{ij}=0.$

It follows from (\ref{C-eq}) and (\ref{C-prop}) that the $\bar{F}$-propagator
is the
heat kernel of the adjoint operator
\be  \left[\mathcal{M}^*\right]^{ij}_{pq}=\delta^i_p\delta^j_q
S^{\alpha\beta}\partial_\alpha\partial_\beta
+\delta^i_p \partial_qS^{\alpha j}\partial_\alpha
+\delta^j_q \partial_pS^{i\beta}\partial_\beta
+\partial_p\partial_q S^{ij}, \lb{Mbar-def} \ee
satisfying
\begin{eqnarray}
\partial_t \bar{F}_{k\ell}^{ij}\left(\brho,0|\br,t\right) & = &
\left[\mathcal{M}^*(\brho)\right]_{k\ell}^{pq}
\bar{F}_{pq}^{ij}\left(\brho,0|\br,t\right) \cr
&=&
\left[\mathcal{M}(\br)\right]_{pq}^{ij}
\bar{F}_{k\ell}^{pq}\left(\brho,0|\br,t\right).
\lb{Fbar-eq} \end{eqnarray}
The propagator $F$ is thus the heat kernel of $\mathcal{M}$:
\begin{eqnarray}
\partial_t F^{k\ell}_{ij}\left(\br,0|\brho,t\right) & = &
\left[\mathcal{M}(\br)\right]^{k\ell}_{pq}F^{pq}_{ij}\left(\br,0|\brho,t\right)
\cr
&=&
\left[\mathcal{M}^*(\brho)\right]^{pq}_{ij}
F^{k\ell}_{pq}\left(\br,0|\brho,t\right)
\lb{F-eq} \end{eqnarray}
Because of the homogeneity of the operators $\mathcal{M}$ and $\mathcal{M}^*$
in the $\nu,\kappa\rightarrow$ limit, $F$ satisfies the scaling relation
\be F^{k\ell}_{ij}\left(\lambda\br,0|\lambda\brho,\lambda^\gamma t\right)
=\lambda^{-d}F^{k\ell}_{ij}\left(\br,0|\brho,t\right), \lb{F-scal} \ee
with an identical relation for the $\bar{F}$-propagator.

Finally, it follows from (\ref{G-prop2}) that the gradient correlation
satisfies
\be \partial_t\mathcal{G}_{ij}(\brho,t)
      = \left[\mathcal{M}^*(\brho)\right]^{pq}_{ij}\mathcal{G}_{pq}(\brho,t).
\lb{G-eq} \ee
This equation has an invariant subspace of solutions of the form
$\mathcal{G}_{ij}
=-\partial_i\partial_j\Theta$ for a scalar correlation function
$\Theta(\br,t).$
 Celani et al. \cite{Celanietal06} have also introduced the quantity
\be \mathcal{Q}_{k\ell}(\br,t)\equiv \int
d^d\rho\,F^{ii}_{k\ell}(\brho,t|\br,0) =
\langle \delta\bell_k(t)\bdot\delta\bell^{\prime}_\ell(t)\rangle, \lb{Q-def}
\ee
where on the righthand side the line-elements are initially unit vectors
$\delta\bell_k(0)=\hat{{\bf e}}_k,\,\delta\bell^{\prime}_\ell(0)=\hat{{\bf
e}}_\ell$
starting at positions displaced by $\br.$ This quantity measures the angular
correlation of the material line-elements at times $t>0,$ as well
as their growth in length. It follows from (\ref{F-eq}) that this quantity
in the KK model satisfies
\be  \partial_t \mathcal{Q}_{k\ell}(\br,t)
      = \left[\mathcal{M}^*(\br)\right]^{pq}_{k\ell}\mathcal{Q}_{pq}(\br,t)
      \lb{Q-eq} \ee
with initial condition $\mathcal{Q}_{k\ell}(\br,0)=\delta_{k\ell},$ as already
noted in \cite{Celanietal06}. This equation is identical to (\ref{G-eq}) for
the
gradient correlation and, furthermore, $\mathcal{Q}_{k\ell}(\br,0)=-\partial_k
\partial_\ell\Theta(\br,0)$ with $\Theta(\br,0)=-(1/2)r^2.$ Thus, $\mathcal{Q}$
is of gradient type.

In this work we restrict ourselves to conditions of statistical homogeneity,
isotropy
and parity invariance for all stochastic quantities. Thus, we can write the
2-point
correlation function of the magnetic field as
\begin{equation}
\mathcal{C}^{ij}=C_L(r,t)\hat{r}^i\hat{r}^j +
C_N(r,t)(\delta^{ij}-\hat{r}^i\hat{r}^j) \end{equation}
where $\hat{r}^i=r^i/r$. $C_L$ and $C_N$ are the longitudinal and transverse
correlations, respectively. With the form of the velocity correlation in
(\ref{vel-corr}),
the evolution equation (\ref{C-eq}) reduces to two coupled equations for
$C_L$ and $C_N.$ A lengthy but straightforward calculation gives
\begin{equation}\label{CL}
\begin{array}{lll}
&&\partial_tC_L\vspace{2mm}\\
&&=D_1r^{\xi} \left\{ \left(d-1\right) \partial_{rr}C_L +
\left(d+1\right)\left(d-\xi-1\right)\frac{1}{r}
\partial_rC_L\right.\vspace{2mm}\\
&&\hspace{4mm}\left.+(d-1)\left[\xi^2-\xi-2(d-1)\right]\frac{1}{r^2}
C_L\right.\vspace{2mm}\\
&&\hspace{4mm}\left.+\left(d-1\right)
\left[\left(d+1\right)\xi+2\left(d-1\right)\right]\frac{1}{r^2}C_N
\right\}.
\end{array}
\end{equation} and
\begin{equation}\label{CN}
\begin{array}{lll}
&&\partial_tC_N\vspace{2mm}\\
&&=D_1r^{\xi} \left\{ \left(d-1\right) \partial_{rr}C_N+
\left[\left(d+1\right)\xi+\left(d-1\right)^2\right]
\frac{1}{r}\partial_rC_N\right.\vspace{2mm}\\
&&\hspace{4mm}+
\left[\left(d+1\right)\xi^2+\left(d^2-5\right)\xi-2\left(d-1\right)\right]
\frac{1}{r^2}C_N\vspace{2mm}\\
&&\left.\hspace{4mm}+\left(\xi-2\right)
\left(\xi-1\right)\left(d+\xi-1\right)\frac{1}{r^2}C_L
\right\},
\end{array}
\end{equation} respectively, when $\nu,\kappa\rightarrow 0.$
For solenoidal solutions, such as for the magnetic field,
it is easy to show that the two correlations are related by
\be C_N=C_L+\frac{1}{d-1}r\partial_rC_L. \lb{C-sol} \ee
For example, see \cite{Saffman63,Batchelor53}. The solutions satisfying this
relation form an invariant subspace, with the evolution reducing to a single
equation
for $C_L$:
\begin{equation}\label{CL-div}
\begin{array}{lll}
\partial_tC_L
&=&D_1r^{\xi} \left[ \left(d-1\right) \partial_{rr}C_L +\left(
2\xi+d^2-1\right)\frac{1}{r}\partial_rC_L\right.\vspace{2mm}\\
&&\left.+(d-1)\xi(d+\xi)\frac{1}{r^2}C_L \right].
\end{array}
\end{equation}

In the same manner, the general solution of (\ref{G-eq}) or (\ref{Q-eq}) may be
decomposed as
\begin{equation}
\mathcal{G}_{ij}=
G_L(r,t)\hat{r}_i\hat{r}_j+G_N(r,t)(\delta_{ij}-\hat{r}_i\hat{r}_j)
\end{equation}
under assumptions of homogeneity, isotropy and reflection-symmetry. Then
$G_N$ and $G_L$ satisfy the following coupled equations for
$\nu,\kappa\rightarrow 0$
\begin{equation}\label{GL}
\begin{array}{lll}
&&\partial_tG_L\vspace{2mm}\\
&&=\left(d-1\right)D_1r^{\xi} \left\{\partial_{rr}G_L+
\left(3\xi+d-1\right)\frac{1}{r}\partial_rG_L\right.\vspace{2mm}\\
&&\hspace{6mm}+\left(3\xi+2d-2\right)\left(\xi-1\right) \frac{1}{r^2}G_L
\vspace{2mm}\\
&&
\hspace{6mm}
+\left(d+\xi-1\right)\left(\xi-1\right)\left(\xi-2\right)\frac{1}{r^2}G_N\big\}
\end{array}
\end{equation}
and
\begin{equation}\label{GN}
\begin{array}{lll}
&&\partial_tG_N\vspace{2mm}\\
&&=D_1r^{\xi} \left\{ \left(d-1\right) \partial_{rr}G_N +\left[
\left(d-3\right)\xi+\left(d-1\right)^2\right]\frac{1}{r}
\partial_rG_N\right.\vspace{2mm}\\
&&\left.\hspace{6mm}+
\left[(d+1)\xi+2(d-1)\right]\frac{1}{r^2}G_L\right.\vspace{2mm}\\
&&\hspace{6mm}+\left[\left(d-1\right)\xi-2\right]\left(\xi+d-1\right)
\frac{1}{r^2}G_N\Big\},
\end{array}
\end{equation}
respectively. Gradient solutions satisfy the constraint
\be G_L=G_N+r\partial_rG_N \lb{G-grad} \ee
where $G_N=-\frac{1}{r}\partial_r\Theta$ in terms of the scalar correlation
function
$\Theta.$ In this invariant subspace of solutions the dynamics reduces to a
single
equation for $G_N$:
\begin{equation}\label{GN-grad}
\begin{array}{lll}
\partial_t G_N
&=& (d-1)D_1r^{\xi} \left[ \partial_{rr}G_N +\left(
2\xi+d+1\right)\frac{1}{r}\partial_rG_N\right.\vspace{2mm}\\
&&\left.\hspace{25mm}+\xi (d+\xi)\frac{1}{r^2}G_N \right].
\end{array}
\end{equation}

\subsection{Line-Vector Correlations}\lb{line}

Kinematic dynamo effect is due ultimately to the stretching of
magnetic field lines as they are passively advected by a chaotic velocity
field.
However, the properties of infinitesimal material line-elements in the KK model
are, at first sight, counterintuitive in this respect.  In order to discuss
stretching
of individual lines, it must be understood that the velocity field is smoothed
at
very small scales $\lesssim\ell_\nu.$ The inertial-range velocity structure
function
in (\ref{vel-corr}) crosses over to a viscous-range form
\begin{equation}
S^{ij}(\br)=D_1 \ell_\nu^{\xi-2}r^2
\left[\left(d+1\right)\delta^{ij}-2\hat{r}^{i}\hat{r}^{j}\right]
\lb{vel-corr-nu}
\end{equation}
for $r\ll\ell_\nu.$ The growth of line-elements in such a smooth velocity
field,
white-noise in time was derived by Kraichnan \cite{Kraichnan74} to be
exponential
\be \langle \delta\ell^2(t)\rangle \approx e^{2 \lambda t} \lb{line-grow} \ee
with the Lyapunov exponent
\begin{eqnarray}
\lambda &= & \frac{1}{d+2}\int_{-\infty}^0 dt\,\Big<\frac{\partial
u_i}{\partial x_j}(\bx,t)
      \frac{\partial u_i}{\partial x_j}(\bx,0)\Big>\cr
      &=& D_1\ell_\nu^{\xi-2} d(d-1).
\lb{Kr74} \end{eqnarray}
See also \cite{LeJan84,LeJan85,Son99}. The ``material''  line-elements of
relevance
to the kinematic dynamo are subject to an additional Brownian motion
proportional to
$\sqrt{2\kappa}$ in (\ref{Eqx(a,t)}). However, this effect of molecular
diffusivity $\kappa$
corresponds just to changing the constant $D_0$ in the Kraichnan-model velocity
covariance (\ref{u-cov}) to $D_0+2\kappa$. Since only velocity-gradients enter
(\ref{Kr74}),
this result for the Lyapunov exponent still holds in the presence of molecular
diffusivity.

It follows from (\ref{Kr74}) that line-stretching is {\it greater} for smaller
$\xi$
and smaller $\nu.$ This may seem a bit perplexing, because the dynamo
{\it fails}  for $\xi$ too small, in the range $0<\xi<\xi_*.$ In that regime,
there
is no dynamo action for $\nu<Pr_c\kappa,$ despite the fact that the stretching
rate becomes larger as $\nu$ decreases. The turbulent kinematic dynamo
cannot be understood as a simple ``competition'' between stretching and
diffusion. What, then,  can account for the presence of the dynamo in the range
$\xi_*<\xi<2$ of smoother velocities in the KK model and failure of the dynamo
in the range $0<\xi<\xi_*$ of very rough velocities? An intriguing suggestion
has been
made by Celani et al. \cite{Celanietal06} that the existence of dynamo effect
can be
characterized by the angular correlations of material line-elements. They
proposed
the function $\mathcal{Q}$ defined in ({\ref{Q-def}) as an ``order parameter''
for the
dynamo transition. As we shall demonstrate below, the principal conclusions
of \cite{Celanietal06} about $\mathcal{Q}$ are erroneous and this quantity
does not discriminate between dynamo and non-dynamo regimes in the KK model.
However, our discussion will lead us to identify a different correlation
property of
infinitesimal material line-vectors, which can  indeed serve as an ``order
parameter''
for the dynamo.

The principal claims of \cite{Celanietal06} were as follows. First, in the
non-dynamo
regime for $0<\xi<\xi_*$ with rough velocity, $\mathcal{Q}$ exhibits a  scaling
law of correlations:
\begin{equation}
Q_{k\ell}(\br,t) \sim
      ({\rm const.})
\left({{r}\over{(D_1t)^{1/\gamma}}}\right)^{\bar{\zeta}}
\bar{Z}_{k\ell}(\hat{\br}),
\lb{celani1}   \end{equation}
for $r\ll \min\{ (D_1t)^{1/\gamma},L\}.$ For finite $\nu,\kappa,$ this relation
holds
in the inertial-convective range of scales $\max\{\ell_\kappa,\ell_\nu\}\ll
r\ll L,$
with $\ell_\kappa=(\kappa/D_1)^{1/\xi}$ (assuming that $\nu< Pr_c\kappa$).
This is an example of  ``zero-mode dominance'' \cite{Bernardetal97}. Thus, the
quantity $\bar{Z}_{k\ell}(\br)$ is a homogeneous zero mode of the operator
${\cal M}^*,$
satisfying $[\mathcal{M}^*]_{k\ell}^{pq}\bar{Z}_{pq}=0,$ with exponent
$\bar{\zeta}(\xi)>0$
for $0<\xi<\xi_*.$ Intriguingly, it was found that $\bar{\zeta}=\zeta+2,$ where
$\zeta$
is the scaling exponent of the zero mode of $\mathcal{M}$ which was shown in
\cite{Vergassola96} to dominate in the magnetic correlation of the KK model for
the same parameter range.  It was furthermore claimed in \cite{Celanietal06}
that
$\bar{\zeta}(\xi_*)=0.$ For $2>\xi>\xi_*,$ on the contrary, it was argued that
$\mathcal{M}^*$
develops point spectrum and that
\be  Q_{k\ell}(\br,t) \sim ({\rm const.}) e^{E_0 t}
\bar{\mathcal{E}}_{k\ell}(\br),
\lb{celani2} \ee
where $E_0$ is the largest positive eigenvalue of $\mathcal{M}^*$ and
$\bar{\mathcal{E}}_{k\ell}(\br)$ is the corresponding eigenfunction. $E_0$ is
numerically equal to the dynamo growth rate. Notice in the limit $Pr\ll 1$
that $E_0\propto 1/t_\kappa=(D_1^2/\kappa^\gamma)^{1/\xi}$
\cite{Vincenzi02}, so that $\kappa$ must be kept nonzero (but with
$Re_m>Re_{m,c}).$
Thus, the ``material lines-vectors'' in $\mathcal{Q}$ are advected by velocity
$\bu$ subject to Brownian noise proportional to $\sqrt{\kappa}.$ The
appropriate
terms proportional to $\kappa$ must then be included in the diffusion
operators $\mathcal{M}$ and $\mathcal{M}^*$ \cite{Celanietal06}.

We shall show that the zero-mode dominance relation (\ref{celani1}) does hold
for
$\mathcal{Q}$ but with a different zero-mode and different scaling exponent
$\bar{\zeta}$
than that claimed by \cite{Celanietal06}.  Furthermore, the scaling relation
(\ref{celani1})
holds over the whole range $0<\xi<2,$ assuming only that
$\max\{\ell_\kappa,\ell_\nu\}
\ll r\ll L,$ with an exponent $\bar{\zeta}(\xi)=-\xi<0$ which does not exhibit
any qualitative
change at the dynamo transition $\xi=\xi_*.$  The exponential growth relation
(\ref{celani2}) does not hold for the quantity $\mathcal{Q}$ anywhere over the
range
$0<\xi<2,$ even if $Pr>Pr_c$ and $Re_m>Re_{m,c}.$

\subsection{Zero-Mode Analysis}

The basic tool of our investigation is a generalization of the {\it slow-mode
expansion} of Bernard et al. \cite{Bernardetal97}. Those authors derived such
an
expansion for the propagator or heat kernel $P(\br,t|\br_0,t_0)$ that describes
the
evolution of a passive scalar in the Kraichnan white-noise velocity ensemble
with
covariance (\ref{u-cov}). However, the derivation of \cite{Bernardetal97} was,
in fact,
axiomatic and applicable to the propagator for any non-positive, self-adjoint
operator,
with absolutely continuous spectrum and homogeneous of degree $-\gamma.$
This derivation showed that the slow mode expansion follows from assumed
meromorphic
properties of the Mellin transform of the propagator and Green's function of
the
operator. In the Appendix of this paper, we generalize this axiomatic
derivation
to the case of the non-Hermitian operators $\mathcal{M}$ and $\mathcal{M}^*,$
in
the non-dynamo regime where both have absolutely continuous spectrum. We refer
the reader to the appendix for details and here just state the essential
results.

The operators $\mathcal{M}$ and $\mathcal{M}^*$ have two types of homogeneous
zero-modes, {\it regular} and {\it singular}. The regular zero modes are
denoted $Z_{(a)}$
and $\bar{Z}^{(a)}$  for $a=1,2,3,...,$ respectively, with exponents $\zeta_a$
and $\bar{\zeta}_a$
whose real parts increase with $a.$ These are ordinary functions which satisfy
the
conditions $\mathcal{M}Z_{(a)}=0$ and $\mathcal{M}^*\bar{Z}^{(a)}=0$ globally.
The singular
zero modes $W_{(a)}$ and $\bar{W}^{(a)}$  for $a=1,2,3,...,$ instead have
exponents $\omega_a$
and $\bar{\omega}_a$ whose real parts decrease with $a$ for $a=1,2,3,...,$
respectively.
These are distributions which satisfy the conditions $\mathcal{M}W_{(a)}=0$ and
$\mathcal{M}^*\bar{W}^{(a)}=0$ only up to contact terms. The scaling exponents
of
the two sets of zero modes are related by
\be  \omega_a+\bar{\zeta}_a^*=-d+\gamma,\,\,\,\,
       \bar{\omega}_a+\zeta_a^*=-d+\gamma \lb{exp-rel} \ee
Above each regular zero mode lies an {\it ascending tower of slow modes}
$Z_{(a,p)}$
and $\bar{Z}^{(a,p)}$ homogeneous of degree $\zeta_{a,p}=\zeta_a+\gamma p$ and
$\bar{\zeta}_{a,p}=\bar{\zeta}_a+\gamma p,$ respectively, satisfying ${\cal
M}Z_{(a,p)}
=Z_{(a,p-1)}$ and ${\cal M}^*\bar{Z}^{(a,p)}=\bar{Z}^{(a,p-1)}$ for
$p=1,2,3,...$ with
$Z_{(a,0)}=Z_{(a)}$ and $\bar{Z}^{(a,0)}=\bar{Z}^{(a)}.$ In addition, below
each singular zero
mode is a {\it descending tunnel of self-similar decay solutions}
$W_{(a,p)}(\br,t)$ and
$\bar{W}^{(a,p)}(\br,t)$ satisfying $\partial_tW_{(a,p)}=\mathcal{M}W_{(a,p)}$
and
$\partial_t\bar{W}^{(a,p)}=\mathcal{M}^*\bar{W}^{(a,p)}$ with
$W_{(a,p)}(\lambda\br,
\lambda^\gamma t)=\lambda^{\omega_a-(p+1)\gamma}W_{(a,p)}(\br,t)$ and
$\bar{W}^{(a,p)}(\lambda\br,\lambda^\gamma
t)=\lambda^{\bar{\omega}_a-(p+1)\gamma}
\bar{W}^{(a,p)}(\br,t).$ These are related to the singular zero modes by ${\cal
M}W_{(a,p-1)}
=W_{(a,p)}$ and ${\cal M}^*\bar{W}^{(a,p-1)}=\bar{W}^{(a,p)}$ for
$p=0,1,2,...$ with
$W_{(a,-1)}(\br,0)=W_{(a)}(\br)$ and
$\bar{W}^{(a,-1)}(\br,0)=\bar{W}^{(a)}(\br).$

In terms of these quantities there are short-distance expansions, for
$\lambda\ll 1,$
both for the $F$-propagator
\begin{equation}
 F^{ij}_{k\ell}(\lambda \br,t|\brho,0)\sim \sum_{a,p\geq 0}
\lambda^{\zeta_a+\gamma p} Z^{ij}_{(a,p)}(\br)
[\bar{W}_{k\ell}^{(a,p)}(\brho,t)]^*,
\lb{F-exp} \end{equation}
and for the $\bar{F}$-propagator
\begin{equation}
 \bar{F}^{ij}_{k\ell}(\lambda \br,t|\brho,0)\sim \sum_{a,p\geq 0}
\lambda^{\bar{\zeta}_a+\gamma p} \bar{Z}_{k\ell}^{(a,p)}(\br)
[W^{ij}_{(a,p)}(\brho,t)]^*.
\lb{Fbar-exp} \end{equation}
See the Appendix for the details of the derivation. Note that these
asymptotic series are generally dominated by their leading terms for $a=1$ and
$p=0,$
corresponding to the regular zero mode with scaling exponent of smallest real
part. Of course, the leading term may give a zero contribution for various
reasons
and then subleading terms will dominate instead. In  order to make use
of this expansion we must calculate explicitly the homogeneous zero modes of
$\mathcal{M}$ and $\mathcal{M}^*.$

To find the isotropic and scale-invariant zero-modes of $\mathcal{M}$
we substitute into (\ref{CL}) and (\ref{CN}) the forms
$$ C_L=A_L r^{\sigma},\,\,\,\,\, C_N = A_N r^{\sigma} $$
giving the matrix equation
$$ \left[\begin{array}{cc}
       M_{LL} & M_{LN} \cr
        M_{NL} & M_{NN} \cr
      \end{array}\right]
       \left[\begin{array}{c}
       A_{L}  \cr
       A_{N} \cr
      \end{array}\right]
      =  \left[\begin{array}{c}
           0 \cr
           0 \cr
      \end{array}\right] $$
with
\begin{eqnarray*}
M_{LL} &= & (d-1)\sigma(\sigma-1)+(d-1)(d-\xi-1)\sigma \cr
              &    & \hspace{5mm} +(d-1)(\xi^2-\xi-2d+2)
\end{eqnarray*}
$$ M_{LN}=(d-1)[(d+1)\xi+2(d-1)]$$
$$M_{NL}=(\xi-2)(\xi-1)(d+\xi-1)$$
\begin{eqnarray*}
M_{NN} &= & (d-1)\sigma(\sigma-1) + [(d+1)\xi+(d-1)^2]\sigma\cr
        &    & +(d+1)\xi^2+(d^2-5)\xi-2d+2.
\end{eqnarray*}
Calculating the determinant
\begin{eqnarray*}
     && \left|\begin{array}{cc}
     M_{LL} & M_{LN} \cr
        M_{NL} & M_{NN} \cr
      \end{array}\right| = (d-1)(\sigma-2) (\sigma+d-2) \times \cr
    &&  \hspace{2mm} \big[ (d-1)\sigma^2+\big(d^2-d+2\xi)\sigma
           + (d-1)\xi(d+\xi)\big],
\end{eqnarray*}
one finds that the scaling exponents $\sigma$ are
\begin{equation*}\label{zeta1}
\zeta_1=-{{d}\over{2}}-{{\xi}\over{d-1}}+{{d}\over{2}}
      \left[1-4\xi{{(d-2)(d+\xi-1)}\over{d(d-1)^2}}\right]^{1/2}
      \end{equation*}
\begin{equation*}\label{omega1}
\omega_2=-{{d}\over{2}}-{{\xi}\over{d-1}}-{{d}\over{2}}
      \left[1-4\xi{{(d-2)(d+\xi-1)}\over{d(d-1)^2}}\right]^{1/2}
      \end{equation*}
$$ \zeta_2 = 2, \,\,\,\, \omega_1 = 2-d. $$
Note that the set $\zeta_1,\omega_2$ correspond to the invariant subspace
of solenoidal solutions,  as may be verified by substituting the scaling
ansatz for $C_L$ into (\ref{CL-div}). The exponent $\zeta_1$ coincides with
that
found by Vergassola \cite{Vergassola96} to dominate in the magnetic 2-point
correlation
for a forced steady-state at high magnetic Reynolds number and zero Prandtl
number.

To find the isotropic and scale-invariant zero-modes of $\mathcal{M}^*$
we likewise substitute into (\ref{GL}) and (\ref{GN}) the forms
$$ G_L=\bar{A}_L r^{\bar{\sigma}},\,\,\,\,\, G_N = \bar{A}_N r^{\bar{\sigma}}
$$
giving the matrix equation
$$ \left[\begin{array}{cc}
       \bar{M}_{LL} & \bar{M}_{LN} \cr
        \bar{M}_{NL} & \bar{M}_{NN} \cr
      \end{array}\right]
       \left[\begin{array}{c}
       \bar{A}_{L}  \cr
        \bar{A}_{N} \cr
      \end{array}\right]
      =  \left[\begin{array}{c}
           0 \cr
           0 \cr
      \end{array}\right] $$
with
\begin{eqnarray*}
\bar{M}_{LL} &= & (d-1)\bar{\sigma}(\bar{\sigma}-1)+(d-1)(3\xi+d-1)\bar{\sigma}
\cr
              &    & \hspace{5mm} +(d-1)(3\xi+2d-2)(\xi-1)
\end{eqnarray*}
$$ \bar{M}_{LN}=(d-1)(d+\xi-1)(\xi-1)(\xi-2)$$
$$\bar{M}_{NL}=(d+1)\xi+2(d-1) $$
\begin{eqnarray*}
\bar{M}_{NN} &= &
(d-1)\bar{\sigma}(\bar{\sigma}-1)+((d-3)\xi+(d-1)^2)\bar{\sigma} \cr
       &    & \hspace{5mm} +((d-1)\xi-2)(d+\xi-1).
\end{eqnarray*}
Calculating the determinant
\begin{eqnarray*}
     && \left|\begin{array}{cc}
     \bar{M}_{LL} & \bar{M}_{LN} \cr
        \bar{M}_{NL} & \bar{M}_{NN} \cr
      \end{array}\right| = (d-1)(\bar{\sigma}+\xi) (\bar{\sigma}+\xi+d) \times
\cr
    &&  \hspace{5mm} \big[
(d-1)\bar{\sigma}^2+\big((d-4)(d-1)+2\xi(d-2)\big)\bar{\sigma} \cr
     && \hspace{10mm}   + 2(d-2)(d+\xi-1)(\xi-1)\big],
\end{eqnarray*}
one finds that the scaling exponents $\bar{\sigma}$ are
$$ \bar{\zeta}_1 = -\xi, \,\,\,\, \bar{\omega}_2 = -(d+\xi)$$
\begin{equation*}\label{zetabar2}
\bar{\zeta}_2={{4-d}\over{2}}+{{2-d}\over{d-1}}\xi+{{d}\over{2}}
      \left[1-4\xi{{(d-2)(d+\xi-1)}\over{d(d-1)^2}}\right]^{1/2}
      \end{equation*}
\begin{equation*}\label{omegabar2}
\bar{\omega}_1={{4-d}\over{2}}+{{2-d}\over{d-1}}\xi-{{d}\over{2}}
      \left[1-4\xi{{(d-2)(d+\xi-1)}\over{d(d-1)^2}}\right]^{1/2}.
      \end{equation*}
Note that the set $\bar{\zeta}_1,\bar{\omega}_2$ corresponds to the invariant
subspace of gradient solutions, as may be verified by substituting the scaling
ansatz for $G_N$ into (\ref{GN-grad}). These exponents are directly
related to those for the passive scalar discussed in \cite{Bernardetal97}.
In the isotropic sector of the scalar, the regular zero mode has exponent
$0$ (constants) and the singular zero mode has exponent $\gamma-d.$
The first vanishes after taking gradients and is replaced by its lowest-lying
slow mode, with exponent $\gamma.$ Taking two space derivatives reduces
these exponents by $2,$ yielding $\gamma-2=-\xi$ and $(\gamma-d)-2=-(d+\xi).$
The exponents $\bar{\zeta}_2,\bar{\omega}_1$ coincide (for $d=3)$ with the set
which were claimed in \cite{Celanietal06} to determine the scaling of
$\mathcal{Q}.$

It is easy to check the relations
\begin{equation*}  \omega_a+\bar{\zeta}_{a}=-d+\gamma,\,\,\,\,
       \bar{\omega}_{a}+\zeta_a=-d+\gamma \lb{exp-rel-KK}
\end{equation*}
for $a=1,2,$
consistent with the general result (\ref{exp-rel}). We see that, at least in
the isotropic
sector, the solenoidal scaling solution $W_{(2)}(\br,t)$ is associated in the
slow-mode
expansion to the non-gradient zero-mode $\bar{Z}^{(2)}(\br)$ and the gradient
scaling
solution $\bar{W}^{(2)}(\br,t)$ is associated to the non-solenoidal zero-mode
$Z_{(2)}(\br).$ In fact, this is true in general, as we now show. Take any
solenoidal scaling solution $W_{(a,p)}(\brho,t).$ Then it follows from  the
propagator relation (\ref{C-prop}) and the scaling property of the singular
slow-mode
$W_{(a,p)}(\br,0)$ that
$$ \int d^dr\,\left[W^{k\ell}_{(a,p)}(\br,0)\right]^*
\bar{F}^{ij}_{k\ell}(\lambda\br,t|\brho,0)
            = \lambda^{\bar{\zeta}_a+\gamma p}
\left[W^{ij}_{(a,p)}(\brho,t)\right]^*. $$
This can only be consistent with the slow-mode expansion (\ref{Fbar-exp}) of
$\bar{F}$ for $\lambda\ll 1,$ if
$$   \int d^dr\,\left[W^{k\ell}_{(a,p)}(\br,0)\right]^*
\bar{Z}_{k\ell}^{(a,p)}(\br)\neq 0. $$
Since  $W_{(a,p)}$  is solenoidal, then $\bar{Z}^{(a,p)}$ must be non-gradient.
Otherwise
the integral will vanish, because the solenoidal and gradient subspaces are
orthogonal.
An identical argument shows likewise that any gradient scaling solution
$\bar{W}^{(a,p)}$
is associated in the slow-mode expansion of  $F$ to a non-solenoidal zero-mode
$Z_{(a,p)}.$

As should now be clear, however, the relation (\ref{celani1}) cannot hold with
$\bar{\zeta}=\bar{\zeta}_2$ and $\bar{Z}=\bar{Z}^{(2)}.$ Since $\mathcal{Q}$
is of gradient type, its evolution is described by (\ref{GN-grad}) which has
$\bar{Z}^{(1)}$ as its only regular zero mode with scaling exponent
$\bar{\zeta}_1.$
We shall now verify this directly from the definition (\ref{Q-def}) of
$\mathcal{Q},$
by means of the slow-mode expansion. We use first the adjoint relation
$F^{ij}_{k\ell}(\brho,t|\br,0)=\bar{F}^{ij}_{k\ell}(\br,t|\brho,0)$ and the
homogeneity
relation (\ref{F-scal}) for $\bar{F}$ to write
$$ \mathcal{Q}_{k\ell}(\br,t) = \int d^d\bar{\rho}\, \bar{F}^{ii}_{k\ell}
     (\lambda\hat{\br},1|\bar{\brho},0) $$
with $\lambda=r/(D_1t)^{1/\gamma}$ and $\bar{\brho}=\brho/(D_1 t)^{1/\gamma}.$
Then using (\ref{Fbar-exp}) for $\lambda\ll 1$ gives
\begin{equation*}
 \mathcal{Q}_{k\ell}(\br,t)\sim \sum_{a,p\geq 0}
\lambda^{\bar{\zeta}_a+\gamma p} \bar{Z}_{k\ell}^{(a,p)}(\hat{\br})
\left[ \int d^d\bar{\rho} \,W^{ii}_{(a,p)}(\bar{\brho},1)\right]^*.
\end{equation*}
Notice, however, that the space integral vanishes for all $W_{(a,p)}$ in the
solenoidal sector (e.g. see \cite{Vincenzi02}). This follows for any solenoidal
correlation $\mathcal{C}$, from
\be \mathcal{C}^{ii}(\br,t)=\partial_k\partial_\ell\mathcal{A}_{k\ell}(\br,t)
              -\triangle\mathcal{A}_{kk}(\br,t), \lb{C-A} \ee
where $\mathcal{A}_{k\ell}$ is the correlation of the vector potential ${\bf
A}.$
Note that, in general dimension $d$, ${\bf B}$ is a 1-form and ${\bf A}$ is a
2-form,
related by the codifferential ${\bf B}=\bdelta{\bf A}$\footnote{Let us give
this argument
in more detail. The representation $\bB=\bdelta\bA$ in dimension $d$ means that
$B^i=\partial_jA^{ij}$ where $A^{ij}=-A^{ji}.$ The relation which replaces
(\ref{C-A})
in general dimension $d$ is
$$ \mathcal{C}^{ij}(\br)=-\partial_k\partial_\ell\mathcal{A}^{ik,j\ell}(\br),$$
where $\mathcal{A}^{ik,j\ell}$ is the 2-point correlation of the 2-form $\bA.$
The result that $\int d^dr\, \mathcal{C}^{ij}(\br)=0$ follows if one assumes
that the correlation function $\mathcal{A}^{ik,j\ell}(\br)\rightarrow 0$
sufficiently
rapidly as $|\br|\rightarrow\infty.$ The result (\ref{C-A}) for $d=3$ is
recovered from
the relation $A^{ij}=\epsilon^{ijk}A_k$ between the 2-form $A^{ij}$ and the
usual
vector potential $A_k.$}.
Since the solenoidal solutions $W_{(a,p)}$ are associated in the expansion
to the slow modes $\bar{Z}^{(a,p)}$ outside the gradient sector, all of these
terms drop out in $\mathcal{Q}.$ The result is the same as the slow-mode
expansion carried out entirely in the gradient sector, with the leading term
\begin{equation}
Q_{k\ell}(\br,t) \sim  C_2
      \left({{r}\over{(D_1t)^{1/\gamma}}}\right)^{\bar{\zeta}_1}
      \bar{Z}_{k\ell}^{(1)}(\hat{\br}),
\lb{celani1-rev}   \end{equation}
for $C_2=\int d^d\rho \, W^{ii}_{(2,0)}(\rho,1).$ This is the correct relation
replacing relation (\ref{celani1}) claimed in \cite{Celanietal06}.

This same relation may be verified by appealing to the results of Eyink and Xin
\cite{EyinkXin00} on the self-similar decay of the passive scalar. Those
authors found that there is a universal form of the self-similar decay
solutions for the scalar correlation function at short distances:
\be \Theta(r,t) \sim \vartheta^2(t)-\frac{\chi(t)}{2\gamma d D_1}r^\gamma
\lb{EyXin} \ee
for $r\ll (D_1t)^{1/\gamma}.$ See \cite{EyinkXin00}, equation (3.21).
Here $\chi(t)=-(1/2)(d/dt)\vartheta^2(t)$ is the scalar dissipation rate.
For general initial data with power-law decay of correlations in space,
$\Theta(r,0)\sim r^{-\alpha},$ the decay rate is given by $\vartheta^2(t)
\sim t^{-\alpha/\gamma}$ at long times \cite{EyinkXin00}.  Since
$\mathcal{Q}_{k\ell}(\br,0)=\delta_{k\ell}$ corresponds to $\Theta(r,0)
=-(1/2)r^2$ with $\alpha=-2,$ we recover from $\mathcal{Q}_{k\ell}(\br,t)
=-\partial_k\partial_\ell\Theta(r,t)$ and eq.(\ref{EyXin}) exactly the relation
(\ref{celani1-rev}), with $\bar{Z}_{k\ell}^{(1)}(\hat{\br})=\delta_{k\ell}-\xi
\hat{\br}_k\hat{\br}_\ell.$ The latter result may be verified from equation
(\ref{G-grad}) by substituting $G_N=\bar{A}_N r^{-\xi}.$ This alternative
derivation of (\ref{celani1-rev}) makes clear its validity over the whole
range $0<\xi<2$ and not just $0<\xi<\xi_*.$  There can be no exponential
growth relation for $\mathcal{Q},$ such as relation (\ref{celani2}) proposed
in \cite{Celanietal06}. It is true that the operator $\mathcal{M}^*$ must
have a positive eigenvalue whenever $\mathcal{M}$ does so. However,
the corresponding eigenfunctions must lie in the non-gradient sector.
An exponential growth for $\mathcal{Q}_{k\ell}(\br,t)=-\partial_k\partial_\ell
\Theta(r,t)$ would require an exponential growth for the scalar correlation
function $\Theta(r,t),$ which does not occur.

\subsection{A Dynamo Order-Parameter}

Based on the previous discussion, we now will propose an alternative
definition of a line-correlation which can serve as an ``order parameter''
for the dynamo transition. Clearly, one should not integrate $F^{ij}_{k\ell}
(\brho,t|\br,0)$ over $\brho,$ as this eliminates the solenoidal sector.
We propose instead to set $\brho=\bzed,$ defining:
\begin{eqnarray}
\mathcal{R}_{k\ell}(\br,t) &=& F^{ii}_{k\ell}(\bzed,t|\br,0) \cr
     &= & \langle
\delta\bell_k(t)\bdot\delta\bell_\ell'(t)\delta^d(\bx(t)-\bx'(t))\rangle,
\lb{R-def} \end{eqnarray}
where, as before, the two line elements are started with
$\delta\bell_k(0)=\hat{{\bf e}}_k,$
$\delta\bell_\ell'(0)=\hat{{\bf e}}_\ell,$ and $\bx'(0)-\bx(0)=\br.$
Because of the delta-function, $\mathcal{R}_{k\ell}(\br,t)$ measures the growth
in
magnitude and the angular correlation between those material line-vectors
which arrive, stretched and rotated,  at the {\it same} point at time $t.$ Just
like
the quantity $\mathcal{Q}$ defined in \cite{Celanietal06}, $\mathcal{R}$
satisfies also the equation
$$ \partial_t \mathcal{R}_{k\ell}(\br,t)
= [\mathcal{M}^*(\br)]^{pq}_{k\ell}\mathcal{R}_{pq}(\br,t). $$
However, it has the initial value
$$\mathcal{R}_{kl}(\br,0)=\delta_{k\ell}\delta^d(\br), $$
which is non-gradient, unlike for $\mathcal{Q}.$ Thus, $\mathcal{R}$ should
experience
exponential growth like (\ref{celani2}) in the dynamo regime for $2>\xi>\xi_*.$

The time-dependence in the non-dynamo regime for $0<\xi<\xi_*$ can be obtained
from the slow-mode expansion of
$\bar{F}^{ii}_{k\ell}(\lambda\hat{\br},1|\bzed,0)$ with
$\lambda=r/(D_1t)^{1/\gamma}.$ One obtains for $\lambda\ll 1$ that
\begin{equation*}
 \mathcal{R}_{k\ell}(\br,t)
 \sim \sum_{a,p\geq 0} (D_1t)^{-\frac{d+\bar{\zeta}_a}{\gamma}-p}
\bar{Z}_{k\ell}^{(a,p)}(\br) \left[W^{ii}_{(a,p)}(\bzed,1)\right]^*
\end{equation*}
Thus, $\mathcal{R}$ exhibits a power-law decay in time, with the dominant terms
 given by the two isotropic zero modes
\begin{eqnarray}
 \mathcal{R}_{k\ell}(\br,t)
 &\sim& C_1 (D_1t)^{-\frac{d+\bar{\zeta}_1}{\gamma}} \bar{Z}_{k\ell}^{(1)}(\br)
\cr
 && \hspace{7mm} +  C_2 (D_1t)^{-\frac{d+\bar{\zeta}_2}{\gamma}}
\bar{Z}_{k\ell}^{(2)}(\br)
\lb{R-scal}
\end{eqnarray}
In all dimensions $d$ the exponent $\bar{\zeta}_2>0$ for $0<\xi<1,$ becoming
negative for $\xi>1.$  Thus, the first term with $\bar{\zeta}_1=-\xi$ dominates
for
lower $\xi$ values. There is a critical dimension $d_c\doteq 4.659,$ given by
the
positive real root of the cubic polynomial $d^3-8d^2+19d-16,$ above which it
instead true that $\bar{\zeta}_2<\bar{\zeta}_1$ when $\xi>\xi_c$ with
$$ \xi_c=\frac{\sqrt{(d^2-3d+4)^2+8(d-1)^2(d-2)}-(d^2-3d+4)}{2(d-1)}. $$
Note that $\xi_c<\xi_*$ where the dynamo transition occurs. The latter value
\cite{ArponenHorvai07}
$$ \xi_* = (d-1)\left(\sqrt{\frac{d-1}{2(d-2)}}-\frac{1}{2}\right) $$
is the point at which $\bar{\zeta}_2$ develops an imaginary part and the
slow-mode
expansion above breaks down.

For exponents $\xi_*<\xi<2$ in all the integer dimensions
$2<d<9,$ the power-law decay is replaced by exponential growth
\be \mathcal{R}_{k\ell}(\br,t)\sim C_0 e^{E_0t}\bar{\mathcal{E}}_{k\ell}(\br).
\lb{R-exp} \ee
proportional to the eigenfunction $\bar{\mathcal{E}}_{k\ell}(\br)$  of
$\mathcal{M}^*$
with largest eigenvalue $E_0.$ To demonstrate this, it is enough to show that
the
initial condition $\mathcal{R}_{k\ell}(\br,0)$ gets a non-zero contribution
from the
eigenfunction $\bar{\mathcal{E}}_{k\ell}(\br).$ We may represent this initial
state
by an expansion
$$
\mathcal{R}_{k\ell}(\br,0)=\sum_{\alpha} C_\alpha
\bar{\mathcal{E}}^\alpha_{k\ell}(\br),
$$
where $\bar{\mathcal{E}}^\alpha_{k\ell}(\br)$ is the eigenfunction of
$\mathcal{M}^*$
with eigenvalue $E_\alpha.$ Note that for the continuous spectrum, this is a
generalized
eigenfunction expansion where the sum over $\alpha$ is a continuous integral
and
$\bar{\mathcal{E}}^\alpha_{k\ell}$ are distributions, not square-integrable
functions.
The expansion coefficients are given by
$$ C_\alpha=\langle \mathcal{E}_\alpha,\mathcal{R}(0)\rangle=
      \int d^dr \, \mathcal{E}^{k\ell}_\alpha(\br) \mathcal{R}_{k\ell}(\br,0)
$$
where $\mathcal{E}^{k\ell}_\alpha$ are the eigenfunctions of $\mathcal{M}$ with
the same eigenvalue $E_\alpha.$ These form a bi-orthogonal set with the
eigenfunctions $\bar{\mathcal{E}}^\alpha_{k\ell}$ of $\mathcal{M}^*$ \footnote{
We remark that related eigenfunction expansions hold for the heat-kernels:
$$ F^{k\ell}_{ij}(\br,0|\brho,t)=\bar{F}^{k\ell}_{ij}(\brho,0|\br,t)
                                             = \sum_\alpha e^{E_\alpha t}
\mathcal{E}^{k\ell}_\alpha(\br)\bar{\mathcal{E}}_{ij}^\alpha(\brho).$$}.
The coefficient $C_0$ corresponding to the eigenfunction
$\bar{\mathcal{E}}^0_{k\ell}=
\bar{\mathcal{E}}_{k\ell}$ is non-zero because
$$ C_0=\int d^dr \, \mathcal{E}^{k\ell}(\br)
\mathcal{R}_{k\ell}(\br,0)=\mathcal{E}^{kk}(\bzed),
$$
where $\mathcal{E}^{kk}(\bzed)\neq 0$ is (twice) the energy in the normalized
dynamo state.

Thus, unlike the quantity $\mathcal{Q}$ proposed in \cite{Celanietal06}, the
line-correlation $\mathcal{R}$ defined in (\ref{R-def}) satisfies the
exponential
growth relation (\ref{R-exp}) in the dynamo regime and power-law scaling
(\ref{R-scal})
in the non-dynamo regime. It would be of great interest to determine the
spatial
structure of the eigenfunction $\bar{\mathcal{E}}_{k\ell}(\br).$ Of course,
this
function must be of non-gradient type.  It is known \cite{Vincenzi02,
BoldyrevCattaneo04} that the trace of the dual eigenfunction $\mathcal{E}(r)=
\mathcal{E}^{ii}(\br)$ exhibits stretched-exponential decay of the form
$\mathcal{E}(r)
\propto -\exp\left(-\beta (r/\ell_\kappa)^{\gamma/2}\right)$ for  $r\gg
\ell_\kappa.$
A similar behaviour for $\bar{\mathcal{E}}(r)=\bar{\mathcal{E}}_{kk}(\br)$ can
be
checked to be consistent with the dynamical equations, but a more careful
investigation
is required. This will be pursued elsewhere.

A quantity with even simpler geometric significance which might also serve
as an ``order parameter'' is
\be \mathcal{R}(t)=\frac{1}{d}\int d^dr\,\mathcal{R}_{kk}(\br,t)
=\langle\delta\bell(t)\bdot\delta\bell'(t)\rangle_0. \lb{R0-def} \ee
This is the covariance of two material line-elements which started at any
relative positions as identical unit vectors at time 0 and which ended at the
{\it same} point at time $t.$  The notation $\langle\cdot\rangle_0$ denotes
the conditional expectation over material lines which end at zero separation.
Clearly $\mathcal{R}(0)=1.$ However, its time-dependence is undetermined
by our present considerations both in the dynamo and in the non-dynamo
regimes.  We cannot argue that $\mathcal{R}(t)$ decays as a power
in the non-dynamo regime, because the slow-mode expansion applies
only for $r\ll(D_1t)^{1/\gamma}$ whereas the definition of $\mathcal{R}(t)$
involves an integral over all $\br.$ We also cannot conclude that
$\mathcal{R}(t)$ grows exponentially in the dynamo regime, because
this requires the condition$\int d^dr\,\bar{\mathcal{E}}_{k\ell}(\br)\neq 0,$
which needs
to be shown. However, we shall see in the next section that $\mathcal{R}(t)$
has a direct interpretation in terms of the turbulent decay of an initially
uniform
magnetic field and we shall determine its time-dependence in the non-dynamo
regime.

\section{Decay of the Magnetic Field}\lb{sec:decay}

We now consider in detail the problem of the turbulent decay of the magnetic
energy $\langle B^2(t)\rangle$ in the non-dynamo regime of the KK model, for
$0<\xi<\xi_*$ and $Pr<Pr_c.$

\subsection{Discussion of the Convective-Range Decay Law}\lb{convect}

We begin by giving a simple, heuristic explanation of the decay law of the
magnetic field, for generic initial data of the magnetic field with rapid
decay of correlations in space. The fundamental observation is that the
zero-modes
$\bar{Z}^{(a)},$ $a=1,2,3,...$ of the adjoint operator $\mathcal{M}^*$ give
rise to
statistical conservation laws in the evolution of the 2-point correlation,
$$ \bar{J}_a(t)\equiv \int d^dr\, \bar{Z}_{ij}^{(a)}(\br)
\mathcal{C}^{ij}(\br,t), $$
which satisfy
\begin{eqnarray*}
(d/dt)\bar{J}_a(t) & = &
\int d^dr\, \bar{Z}_{ij}^{(a)}(\br) \cdot
[\mathcal{M}(\br)]^{ij}_{pq}\mathcal{C}^{pq}(\br,t) \cr
& = & \int d^dr\, [\mathcal{M}^*(\br)]^{ij}_{pq}\bar{Z}_{ij}^{(a)}(\br) \cdot
\mathcal{C}^{pq}(\br,t)
=0.
\end{eqnarray*}
Note, however, that only the {\it non-gradient} zero modes lead to non-trivial
conservation laws, because of the orthogonality of solenoidal and gradient
correlations. The leading-order zero-mode is thus $\bar{Z}^{(2)}$ found in
the previous section, namely,
$$ \bar{Z}^{(2)}_{ij}(\br) = r^{\bar{\zeta}_2}\left[
\bar{A}_L^{(2)}\hat{r}_i\hat{r}_j
     + \bar{A}_N^{(2)} (\delta_{ij}-\hat{r}_i\hat{r}_j)\right], $$
with
$$\bar{A}_L^{(2)} = (\xi-2)[(d-1)\bar{\zeta}_2+(d-3)(d+\xi-1)] $$
$$ \bar{A}_N^{(2)} = (d+1)\xi+2(d-1). $$
This zero-mode coincides with that found by Celani et al. \cite{Celanietal06}
for $d=3.$

The corresponding conserved quantity $\bar{J}_2(t)$ plays the same role
in the turbulent decay of the magnetic field as played by the ``Corrsin
invariant'' in the decay of the passive scalar \cite{EyinkXin00,Chavesetal01}.
Assume, in fact, a self-similar decay law for the magnetic correlation
$$ \mathcal{C}^{ij}(\br,t) = \mathpzc{h}^2(t) \Gamma^{ij}(\br/L(t)). $$
The length $L(t)$ is a large-distance correlation length or ``integral
length'' of the magnetic field.  The quantity $\mathpzc{h}(t)$ is a measure
of the magnitude of the magnetic fluctuations at scale $L(t),$ which we
term the {\it magnetic amplitude}. Just as for the scalar, the growth of the
magnetic
length-scale $L(t)$ can be obtained dimensionally from
\be \frac{1}{L(t)}\frac{d}{dt}L(t)= D_1 L^{-\gamma}(t), \lb{L-eq} \ee
yielding
\be   L(t) = [L^\gamma(0)+ \gamma D_1(t-t_0)]^{1/\gamma}. \lb{length} \ee
To determine the decay rate requires a relation between $\mathpzc{h}(t)$ and
$L(t)$ which is provided by invariance of $\bar{J}_2$:
\begin{equation}\label{JB}
\bar{J}_2=\mathpzc{h}^2(t) L^{d+\bar{\zeta}_2}(t)C
\end{equation}
with $C=\int d^d\rho\,\bar{Z}_{ij}^{(2)}(\brho)\Gamma^{ij}(\brho).$ Thus,
finally,
\be \mathpzc{h}^2(t) \sim \bar{J}_2 [L(t)]^{-(d+\bar{\zeta}_2)}
      \sim (t-t_0)^{-(d+\bar{\zeta}_2)/\gamma}
\lb{decay} \ee
for $(t-t_0)\gg L^\gamma(0)/D_1.$ The generic decay of the magnetic amplitude
is predicted by this argument to be determined by the scaling exponent
$\bar{\zeta}_2,$
which decreases with increasing $\xi$ over the range $0<\xi<\xi_*.$  Thus,
the decay rate is faster for rougher velocities and slower for smoother
velocities.
It is noteworthy that the decay law (\ref{decay}) is completely independent
of the resistivity.

The above argument does not apply if $\bar{J}_2=0.$ In that case, one can
expect that invariants
$$ \bar{J}_{a,p}(t)\equiv \int d^dr\, \bar{Z}_{ij}^{(a,p)}(\br)
\mathcal{C}^{ij}(\br,t), $$
from higher-order zero modes and slow modes of $\mathcal{M}^*$ (again in
the non-gradient sector) will determine the decay rate . Note, for example,
that
$(d/dt)\bar{J}_{2,1}(t)=\bar{J}_{2}(t),$ so that $\bar{J}_{2,1}$ becomes
invariant
if  $\bar{J}_2=0.$ In ref.\cite{EyinkXin00} it was shown that there are two
universality classes in the turbulent decay of the passive scalar for generic
initial
data with rapidly decaying correlations in space, depending upon whether
the ``Corrsin invariant'' $J_0$ from the constant zero mode is vanishing or
nonvanishing. If $J_0=0$, then there exists a higher-order invariant
$J_1\neq0,$
associated to the first slow mode $r^\gamma$ in the tower above the constant
zero
mode, which determines the decay. Chaves et al. \cite{Chavesetal01} showed
how this picture emerges from the slow-mode expansion of Bernard et al.
\cite{Bernardetal97} and extends to the higher-order correlations of the
scalar.
In the following section we shall present a similar treatment of the turbulent
decay
of the magnetic field, based on our generalized slow-mode expansion in the
Appendix.

There are essential differences, however, between the turbulent decay of
a passive scalar and of a passive magnetic field.  Whereas the scalar field
has a finite limit as diffusivity $\kappa\rightarrow 0,$ this is {\it not} true
for
the magnetic field which, even in the non-dynamo regime, tends to accumulate
at the resistive scale \cite{Vergassola96}. As we shall see below, the scaling
function $\Gamma^{ij}(\brho)$ grows with $\rho$ decreasing through the
convective range. Thus, one cannot set $\rho=0$ to interpret $\mathpzc{h}^2(t)$
as the magnetic energy. A more correct interpretation of the magnetic
amplitude is that $\mathpzc{h}^2(t)/L^2(t)\simeq \langle |\bA(t)|^2\rangle,$
where $\bA$ is a vector potential (2-form) such that $\bB=\bdelta\bA.$
The decay rate of the magnetic energy $\langle |\bB(t)|^2\rangle$ cannot
be obtained from purely ideal considerations, but requires an explicit matching
of
convective-range solutions with resistive-scale solutions. In the following two
sections we treat first the ideal, convective range problem with
$\kappa\rightarrow 0.$

\subsection{Self-Similar Decay for Initial Data with Short-Range
Correlations}\lb{short}

Consider any initial 2-point correlation function $\mathcal{C}^{ij}(\br,0)$ of
the magnetic
field which decreases rapidly for large $r.$ We shall demonstrate that the
correlation
$\mathcal{C}^{ij}(\br,t)$ at much later times exhibits self-similar decay and
determine
the decay law.  We use the propagator relation (\ref{C-prop}) and the symmetry
properties of $\bar{F}$ to write:
\begin{eqnarray}
\mathcal{C}^{ij}(\br,t) &=&\displaystyle\int d^d\rho\,
\mathcal{C}^{k\ell}(\brho,0)\bar{F}^{ij}_{k\ell}(\brho,t|\br,0) \cr
&=& \lambda^d \displaystyle\int d^d\rho\,
\mathcal{C}^{k\ell}(\brho,0)\bar{F}^{ij}_{k\ell}(\lambda\brho,1|\bar{\br},0)
\end{eqnarray}
with $\lambda=1/(D_1t)^{1/\gamma}$ and $\bar{\br}=\br/(D_1t)^{1/\gamma}.$
In the last line we used the scaling property (\ref{F-scal}) for $\bar{F}.$
Since $\mathcal{C}^{k\ell}(\brho,0)$ decays rapidly for $\rho\gg L(0),$ we
may employ the slow-mode expansion (\ref{Fbar-exp}) for $(D_1t)^{1/\gamma}\gg
L(0).$ Because $\mathcal{C}^{k\ell}(\brho,0)$ is solenoidal, only the
non-gradient
zero-modes of $\mathcal{M}^*$ give a non-vanishing contribution.

The leading-order term, in general, is
\begin{eqnarray}
\mathcal{C}^{ij}\left(\br,t\right) &\sim&(D_1t)^{-(d+\bar{\zeta}_2)/\gamma}
\left(\displaystyle\int d^d \rho\, \mathcal{C}^{k\ell}(\brho,0)
\bar{Z}_{k\ell}^{(2)}(\brho)\right)\cr
&& \hspace{10mm} \times
W^{ij}_{(1)}\left(\displaystyle\frac{\br}{\left(D_1t\right)^{1/\gamma}},
\right)
\vspace{2mm}\\
&\sim& \left(\displaystyle\int d^d \rho \,
\mathcal{C}^{k\ell}(\brho,0)\bar{Z}_{k\ell}^{(2)}(\brho)\right)
W^{ij}_{(1)}\left(\br,t\right). \cr
\nonumber \lb{C-longT} \end{eqnarray}
In the last line we have used the self-similarity property
$$W_{(1)}(\lambda\br,
\lambda^\gamma t)=\lambda^{-(d+\bar{\zeta}_2)}W_{(1)}(\br,t).$$
We have also used the fact that $W_{(1)}$ is a real-valued function.  This
will be demonstrated in the following section, where we shall derive the
explicit
functional form of all the self-similar decay solutions. We conclude that, as
long
as $\bar{J}_2(0)\neq 0,$ then the generic magnetic correlation
$\mathcal{C}^{ij}(\br,t)$
with short-range initial data is proportional at long times to the self-similar
decay
solution $W_{(1)}^{ij}(\br,t).$

It is important to demonstrate that the above scenario is statistically
realizable
\cite{EyinkXin00}. We shall construct now a positive-definite covariance
function
for which $\bar{J}_2(0)\neq 0.$  This will also demonstrate the
positive-definiteness
of the scaling solution $W_{(1)},$ since the dynamics is
realizability-preserving and
the above argument shows that
$$ \lim_{\lambda\rightarrow\infty}
\lambda^{d+\bar{\zeta}_2}\mathcal{C}(\lambda\br,
\lambda^\gamma t)=\bar{J}_2(0)\cdot W_{(1)}(\br,t). $$
As a simple example we take, with $\mathcal{N}=(\sigma/\sqrt{2\pi})^d,$
$$\begin{array}{lll} C^{ij}(\brho,0)&=& \mathcal{N}\int
d^dk\left(k^2\delta^{ij}-k^ik^j\right)\exp\left(-\frac{\sigma^2k^2}{2}\right)
e^{i \brho\bdot\bk}\vspace{2mm}\\
&=&\left(-\triangle_{\rho}\delta^{ij}
+\partial_{\rho}^{i}\partial_{\rho}^{j}\right)
\exp\left(-\frac{\rho^2}{2\sigma^2}\right).
\end{array}$$
A bit of calculation shows for this example that
\begin{eqnarray*}
\bar{J}_2(0) &= & (d-1)(2\sigma^2)^{(d+\bar{\zeta}_2-2)/2}S_{d-1}
     \Gamma\left(\frac{d+\bar{\zeta}_2}{2}\right) \cr
&& \hspace{1cm}\times
     \left[ \bar{A}_L^{(2)}-(\bar{\zeta}_2+1)\bar{A}_N^{(2)}\right],
\end{eqnarray*}
where $S_{d-1}=2\pi^{d/2}/\Gamma\left(\frac{d}{2}\right)$ is the hypersurface
area
of the unit sphere in $d$-dimensions and $\bar{A}_L^{(2)},\bar{A}_N^{(2)}$ are
the
coefficients given in the previous section. At generic values of $d$ and $\xi,$
$\bar{J}_2(0)\neq 0.$ It is noteworthy that $\bar{J}_2(0)=0$ in the example
above
precisely at the point of degeneracy of zero-modes where
$\bar{\zeta}_1=\bar{\zeta}_2.$
As discussed in section \ref{line}, this occurs for $d>d_c\doteq 4.659$ at the
single
value $\xi=\xi_c<\xi_*.$ In fact, $\bar{J}_2(0)=0$ at this point for all
initial data,
because there is then a single zero-mode of gradient-type satisfying $
\bar{A}_L^{(2)}
=(\bar{\zeta}_2+1)\bar{A}_N^{(2)}$; see (\ref{G-grad}). Of course, whenever
$\bar{J}_2(0)
=0$ then higher-order terms in the slow-mode expansion become dominant and a
different self-similar solution $W_{(a,p)}(\br,t)$ becomes the long-time
attractor.
We shall defer to future work the study of this non-generic situation.



In the remainder of this section we shall make some important comments
about the generic case $\bar{J}_2(0)\neq 0.$ Our first observation is about
the property of  ``quasi-equilibrium''. It was shown in
\cite{EyinkXin00,Chavesetal01}
that the short-distance scaling of the scalar structure function in the decay
of the
passive scalar is identical to the scaling of the scalar structure function in
a
forced steady-state. This is the property of turbulence decay traditionally
termed
 ``quasi-equilibrium.'' We show here a similar property for the turbulent decay
of
 magnetic field, using the slow-mode expansion, as in \cite{Chavesetal01} for
 the scalar. We use the propagator $F,$ its scaling property (\ref{F-scal}),
 and the change of variables
 $\bar{\brho}=\lambda\brho$ with $\lambda=1/(D_1t)^{1/\gamma}$ to write
 \begin{eqnarray*}
\mathcal{C}^{ij}(\br,t) &=&\displaystyle\int d^d\rho\,\,
\mathcal{C}^{k\ell}(\brho,0) F^{ij}_{k\ell}(\br,t|\brho,0) \cr
&=& \displaystyle\int
d^d\bar{\rho}\,\,\mathcal{C}^{k\ell}(\bar{\brho}/\lambda,0)
F^{ij}_{k\ell}(\lambda\br,1|\bar{\brho},0). \end{eqnarray*}
We now employ the slow-mode expansion (\ref{F-exp}) of $F$ for $r\ll
(D_1t)^{1/\gamma}$
to conclude that
\begin{eqnarray}
\mathcal{C}^{ij}(\br,t) &= &
\left(\displaystyle\frac{r}{(D_1t)^{1/\gamma}}\right)^{\zeta_1}
Z^{ij}_{(1)}(\hat{\br}) \\
&& \times \hspace{2mm} \int d^d\bar{\rho} \,\,
 C^{k\ell}\left((D_1t)^{1/\gamma}\bar{\brho},0\right)
\bar{W}_{k\ell}^{(2)*}\left(\bar{\brho},1\right). \cr
\nonumber \lb{C-smallR} \end{eqnarray}
The scaling exponent $\zeta_1$ and zero-mode $Z_{(1)}$ are the same as
found in \cite{Vergassola96} to determine the short-distance scaling of the
magnetic
correlation function in the forced steady-state, which is just the
``quasi-equilibrium''
property \footnote{Note that
(57)
applies for small $r$ at fixed times $t,$ whereas
(56)
applies at long times $t$
for fixed $\br.$ However, the two results agree in their common domain of
validity
for $r,L(0)\ll (D_1 t)^{1/\gamma}.$ This may be seen by applying
(57)
to $W_{(1)}$ to obtain for $r\ll (D_1 t)^{1/\gamma}$
$$ W^{ij}_{(1)}\left(\frac{\br}{\left(D_1t\right)^{1/\gamma}},1\right)
      \sim C \left(\displaystyle\frac{r}{(D_1t)^{1/\gamma}}\right)^{\zeta_1}
Z^{ij}_{(1)}(\hat{\br}). $$
This result is verified in Section \ref{similar} with the explicit expression
for
$W_{(1)}.$ Substituting the above into (\ref{C-longT}) gives
$$ \mathcal{C}^{ij}\left(\br,t\right) \sim C
(D_1t)^{-(d+\zeta_1+\bar{\zeta}_2)/\gamma}
     Z^{ij}_{(1)}(\br). $$
This same result may be obtained by changing the integration variable in
(57)
back to $\brho=\bar{\brho}/(D_1 t)^{1/\gamma}$ and then
employing the similar ``quasi-equilibrium'' result
$$
\bar{W}_{k\ell}^{(2)}\left(\frac{\brho}{\left(D_1t\right)^{1/\gamma}},1\right)
      \sim C
\left(\displaystyle\frac{\rho}{(D_1t)^{1/\gamma}}\right)^{\bar{\zeta}_2}
\bar{Z}_{k\ell}^{(2)}(\hat{\brho}).$$
substituted into
(57).
}.
Since $\zeta_1<0$ for all $0<\xi<\xi_*,$
we see that $\mathcal{C}^{ij}(\br,t)$ increases without bound as $r$ decreases,
in agreement with our earlier physical discussion. We shall confirm this
result by an independent argument in the next section.


A second observation is that the above discussion---the demonstration
of self-similar decay and quasi-equilibrium---carry over directly to the
general $n$-point correlation function of the magnetic field. Note that
$$
\begin{array}{lll}
&& C_n^{i_1i_2\ldots
i_n}(\br,t)=\left<B^{i_1}(\bx_1,t)B^{i_2}(\bx_2,t)\ldots
B^{i_n}(\bx_n,t)\right>\vspace{2mm}\\
&&\hspace{1cm}=\displaystyle\int d^{d}\rho \,\,C_n^{j_1j_2\ldots j_n} (\brho,0)
\bar{F}^{i_1i_2\ldots i_n}_{n,j_1j_2\ldots j_n}(\brho,0 |\br,t)
\end{array}
$$
where $\br=(\bx_1,\bx_2,\ldots \bx_n)$ and $F_n$ is the n-body propagator.
All the symmetries used in the previous argument hold for general $n$,
e.g., time-reversal and
$F_n\left(\brho,t\,\vrule\,\br,0\right)=\bar{F}_n\left(
\br,0\,\vrule\,\brho,t\right)$. Note that due to space homogeneity only
the separation of the variables matter (the absolute position of each particle
is irrelevant)
and we can work in the $(n-1)d$-dimensional separation-of-variables sector.
$F_n$ then
has the scaling property
$$
F_n\left(\lambda\brho,0\,\vrule\,\lambda
\br,\lambda^{\gamma}t\right)=\lambda^{-d(n-1)}F_n\left(\brho,0|\br,t\right). $$
Finally, slow-mode expansions like (\ref{F-exp}) and (\ref{Fbar-exp}) are valid
for $F_n$ and $\bar{F}_n$ for all integers $n.$ See \cite{Bernardetal97} and
the Appendix for details.  The whole analysis thus goes through as
for $n=2$ above and as in \cite{Chavesetal01} for the scalar case.



\subsection{General Self-Similar Decay}\lb{similar}




To complement the previous discussion employing the slow-mode expansion
we shall here determine all possible self-similar decay solutions for the
magnetic correlation function $\mathcal{C},$ following the analysis in
\cite{EyinkXin00} for the passive scalar. It is convenient to employ the
longitudinal correlation $C_L$ which satisfies the equation (\ref{CL-div}).
We introduce the self-similar ansatz
\begin{equation}\label{self-similar}
C_L(r,t)=\mathpzc{h}^2(t)\Gamma\left(\frac{r}{L(t)}\right).
\end{equation}
Substituting the ansatz (\ref{self-similar}) into equation (\ref{CL-div}) for
$C_L$
we arrive at, with $\rho=r/L$,
\begin{equation}\label{Eqrhovariable}
\begin{array}{lll}
&&\displaystyle \frac{1}{D_1L^{-\gamma}(t)}
\frac{2\dot{\mathpzc{h}}(t)}{\mathpzc{h}(t)}\Gamma(\rho)-
\displaystyle\frac{1}{D_1L^{-\gamma}(t)}
\frac{\dot{L}(t)}{L(t)}\rho\Gamma'(\rho)\vspace{2mm}\\
&&=(d-1)\rho^{\xi}\Gamma''(\rho)
+(2\xi+d^2-1)\rho^{\xi-1}\Gamma'(\rho)
\vspace{2mm}\\
&& \hspace{1cm}+\xi(d-1)(d+\xi)\rho^{\xi-2}\Gamma(\rho)
\end{array}\end{equation}
This implies that
\begin{equation}\label{h}
\frac{2\dot{\mathpzc{h}}(t)}{\mathpzc{h}(t)}= -\alpha D_1 L^{-\gamma}(t)
\end{equation}
\begin{equation}\label{L}
\frac{\dot{L}(t)}{L(t)}=\beta D_1L^{-\gamma}(t).
\end{equation}
with constants $\alpha$ and $\beta.$
We have freedom in choosing the value of $\beta$ to fix the length-scale;
here we adopt $\beta=1$. The equation for $L(t)$ then becomes identical
to (\ref{L-eq}) with solution (\ref{length}). Combining (\ref{h}) and (\ref{L})
yields $2\dot{\mathpzc{h}}(t)/\mathpzc{h}(t)=-\alpha\dot{L}(t)/L(t)$ with
solution
\be \mathpzc{h}^2(t)=[L(t)]^{-\alpha}. \lb{hpzc} \ee

Employing (\ref{h}) and (\ref{L}), the equation (\ref{Eqrhovariable}) for the
scaling
function $\Gamma$ becomes
\begin{equation}\label{Hcal-eq}
\begin{array}{lll}
&&\displaystyle \rho^\gamma\left[\rho\Gamma'(\rho)
+\alpha\Gamma(\rho)\right]\vspace{2mm}\\
&&=(d-1)\rho^{2}\Gamma''(\rho)
+(2\xi+d^2-1)\rho\Gamma'(\rho)
\vspace{2mm}\\
&& \hspace{1cm}+\xi(d-1)(d+\xi)\Gamma(\rho)
\end{array}\end{equation}
Making the substitution $x=-\rho^\gamma/\gamma(d-1)$ yields
\begin{eqnarray*}
&& \gamma^2x^2\Gamma_{xx}+
\left[\gamma\left(d+\gamma+\frac{2\xi}{d-1}\right)-\gamma^2x\right]x\Gamma_x
\cr
&& \hspace{1cm}    +\left[\xi(d+\xi)-\frac{\alpha\gamma}{d-1}\right]\Gamma=0.
\end{eqnarray*}
An equation of this form can be solved by the Frobenius method (e.g. see
\cite{Teschl09},
Sec.4.2). According to the general theory, there are two independent solutions
of
the form $\Gamma(x)=x^b\Phi(x),$ where $b$ is a root of the indicial equation
$$ \gamma^2b(b-1)+\left(d+\gamma+\frac{2\xi}{d-1}\right)\gamma b+\xi(d+\xi)=0.
$$
If the two roots are distinct and do not differ by an integer, then the two
functions
$\Phi$ are both analytic, given by convergent power series. Otherwise, only one
solution must be analytic and the second may be an analytic function plus $C\ln
x$
times the first. In our case, it is easy to check that the roots of the
indicial equation
are just given by $b=\zeta_1/\gamma,\omega_2/\gamma$ in terms of the scaling
exponents of the zero-modes of $\mathcal{M}.$ If we substitute $\Gamma=
x^{\zeta_1/\gamma}\Phi$ into the equation for $\Gamma,$ we obtain the Kummer
equation \cite{Erdelyi53}
\begin{equation}\label{Kummer}
x\Phi_{xx}+(c-x)\Phi_x-a\Phi=0.
\end{equation}
with
$$ a= \frac{\alpha+\zeta_1}{\gamma}, \,\,\,\,\,
      c= \frac{1}{\gamma}\left(2\zeta_1+d+\gamma+\frac{2\xi}{d-1}\right). $$
Both independent solutions can be obtained from this equation. The first is
the Kummer function $\Phi(a,c;x),$ an entire function given by the power series
\begin{equation}
\Phi(a,c;x)=\sum_{n=0}^{\infty}\frac{(a)_n}{(c)_n}\frac{x^n}{n!}
\end{equation}
with $(a)_n=a(a+1)\ldots(a+n-1).$ The other is the Kummer function
of the second kind, $\Psi(a,c;x),$ which is defined by a suitable linear
combination of $\Phi(a,c;x)$ and $x^{1-c}\Phi(a-c+1,2-c;x).$
See \cite{Erdelyi53}, 6.5.6. It is not hard to check that this second term
corresponds to the root $b=\omega_2/\gamma$ of the indicial equation.
However, we can argue as in \cite{Vergassola96} that matching the solutions
in the convective range with those in dissipation range permits only the
regular zero
mode as an admissible physical solution. Thus, we obtain $\Gamma(x)=
x^{\zeta_1/\gamma}\Phi(a,c;x).$

This result can be simplified somewhat by appealing to the relation
$$ \bar{\zeta}_2= \zeta_1 +\gamma + \frac{2\xi}{d-1}, $$
which follows by combining $\zeta_1+\omega_2=-d-2\xi/(d-1)$ and
$\omega_2+\bar{\zeta}_2=-d+\gamma.$ Note that the above relation
generalizes the result $\bar{\zeta}_2=\zeta_1+2$ for $d=3$ found in
\cite{Celanietal06}. With this relation we obtain $c=(\zeta_1+\bar{\zeta}_2+d)
/\gamma,$ so that
\begin{equation}\lb{Hcal}
\Gamma(\rho)=\rho^{\zeta_1}
\Phi\left(\frac{\alpha+\zeta_1}{\gamma},\frac{\zeta_1+\bar{\zeta}_2+d}{\gamma};
-\frac{\rho^{\gamma}}{(d-1)\gamma}\right).
\end{equation}
All the self-similar solutions of Eq.(\ref{CL-div})  are given by the ansatz
(\ref{self-similar})
with a scaling function of the form in (\ref{Hcal}) above and with $L(t)$ and
$\mathpzc{h}(t)$
given by Eqs.(\ref{length}) and (\ref{hpzc}), respectively. Since $\Phi(0)=1$
is finite,
all of these self-similar solutions satisfy the condition of
``quasi-equilibrium," showing
the same scaling $r^{\zeta_1}$ for $r\ll L(t)$ as found in \cite{Vergassola96}
for the
forced steady-state.

There are two distinct types of self-similar decay solutions corresponding to
different choices of the free parameter $\alpha.$ When $\alpha=\bar{\zeta}_2
+d+p\gamma,$ for $p=0,1,2,\ldots$, then $a=c+p$ with $p=0,1,2,\ldots$. In this
case
$$
\Phi(c+p,c;-x) =\frac{p!}{(c)_{p}} L^{c-1}_{p}(x) e^{-x}.
$$
where $L^{c-1}_{p}(x)$ is the generalized Laguerre polynomial of degree $p$.
(See \cite{Erdelyi53}, Ch.6). This series of solutions has
stretched-exponential
decay in space. If, for example, we take $\alpha=\bar{\zeta}_2+d$ corresponding
to $p=0,$ then we get
$$
\Gamma(\rho)=\rho^{\zeta_1}\exp\left(-\frac{1}{d-1}\frac{\rho^{\gamma}}{\gamma}
\right). $$
The corresponding self-similar decay solution satisfies
$\mathcal{C}(\lambda\br,
\lambda^\gamma t)=\lambda^{-(d+\bar{\zeta}_2)}\mathcal{C}(\br,t).$ The $\alpha=
\bar{\zeta}_2+d$ solution thus coincides with the self-similar solution
$W_{(1)}(\br,t)$
which was shown in the previous section to describe the long-time decay of
generic
initial data with short-range correlations. More generally, the solutions with
$\alpha
=\bar{\zeta}_2+d+\gamma p$ coincide with the self-similar solutions
$W_{(1,p)}(\br,t)$
for $p=0,1,2,\ldots$ which appear in the slow-mode expansion (\ref{Fbar-exp})
of
the adjoint propagator $\bar{F}.$

For any other choice of $\alpha\neq \bar{\zeta}_2+d+\gamma p$ with
$p=0,1,2,\ldots$
one obtains instead a class of self-similar decay solutions with power-law
decay of
correlations at large distances. This follows from the asymptotic relation
$ \Phi(a,c;-x) \sim \frac{\Gamma(c)}{\Gamma(c-a)} x^{-a} $
for ${\rm Re}\,x\rightarrow +\infty,$ if $a\neq c+p,\,p=0,1,2,\ldots
$(\cite{Erdelyi53}, 6.13.1).
Using the above relation together with (\ref{Hcal}), (\ref{self-similar}), and
(\ref{hpzc})
gives for  any self-similar  solution with $\alpha\neq \bar{\zeta}_2+d+\gamma
p,\,\,\,
p=0,1,2,\ldots,$
$$ C_L(r,t) \sim A r^{-\alpha},\hspace{1cm}  r\gg L(t), $$
where $A$ is a {\it time-independent} constant.  This result is usually called
the
``permanence of the large-scale eddies'' in the turbulence literature. Note
that
for initial data with such power-law decay of correlations, the relation
between
$\mathpzc{h}(t)$ and $L(t)$ that determines the decay rate is obtained from
this
permanence, as $\mathpzc{h}^2(t)\simeq A [L(t)]^{-\alpha},$ in agreement
with (\ref{hpzc}). See \cite{EyinkXin00} for more discussion.

\subsection{Decay Law of the Magnetic Energy}\lb{energy}

We are now ready to discuss the decay law for the magnetic energy:
$$ E(t) = \frac{1}{2}\langle |\bB(t)|^2\rangle= {\rm tr}\,\mathcal{C}(\bzed,t).
$$
Under the assumption of isotropic statistics made here, $E(t)=(d/2)C_L(0,t).$
Clearly, in order to evaluate this expression at $r=0,$ we must consider the
matching of our convective range solution to the resistive scales. We may
do this heuristically, as follows.  We assume that, to leading order,
$$ E(t) \simeq (d/2)C_L(\ell_\kappa,t). $$
We then estimate the correlation function on the right by matching with the
convective-range expression
$$ C_L(r,t)\simeq C_0 \mathpzc{h}^2(t)\left(\frac{r}{L(t)}\right)^{\zeta_1}
                    \simeq C_0 [L(t)]^{-(\alpha+\zeta_1)} r^{\zeta_1} $$
for $r\ll L(t)$ and some positive constant $C_0.$ This yields
$$ E(t) \simeq C_1 [L(t)]^{-(\alpha+\zeta_1)} \ell_\kappa^{\zeta_1} $$
for $C_1=(d/2)C_0.$ Although the energy magnitude increases as resistivity
is lowered, the decay rate is independent of resistivity. Since $\zeta_1<0,$
we see that the decay of magnetic energy $E(t)\propto
(t-t_0)^{-(\alpha+\zeta_1)/
\gamma}$ is always slower than the decay of magnetic amplitude
$\mathpzc{h}^2(t)\propto (t-t_0)^{-\alpha/\gamma}.$ For example, for the
generic case with $\alpha=\bar{\zeta}_2+d$ we obtain that
$$ E(t) \propto (t-t_0)^{-c}, $$
with $c=(\zeta_1+\bar{\zeta}_2+d)/\gamma.$

The above argument is basically correct, but not fully rigorous. It seems
worthwhile to give a more systematic derivation, since the time-dependence
of magnetic energy is crucial to the issue of whether dynamo action is present
or not. We employ a standard method of matched asymptotic expansions (see,
for instance, Chap. V of Ref. \cite{VanDyke64}).
The equation obeyed by $C_L$ for $\kappa>0$ is
\begin{eqnarray}
\partial_t C_L &=&\left[(d-1)r^{\xi}\partial_r^2C_L
+(2\xi+d^2-1)r^{\xi-1}\partial_r C_L\right. \cr
&&\hspace{1cm}\left.+\xi(d-1)(d+\xi)r^{\xi-2}C_L\right] \cr
&&\hspace{15mm}+2\kappa
\left[\partial_r^2C_L+(d+1)\frac{1}{r}\partial_rC_L\right].
\end{eqnarray}
See \cite{ArponenHorvai07}.  Substituting the self-similar ansatz
(\ref{self-similar})
we get
\begin{equation}\label{outerscaling}
\begin{array}{lll}
&&\left[\alpha\Gamma+\rho \Gamma_{\rho}\right]
+\left[(d-1)\rho^{\xi}\Gamma_{\rho\rho}+
(2\xi+d^2-1)\rho^{\xi-1} \Gamma_{\rho}\right.\vspace{2mm}\\
&&\left.+\xi(d-1)(d+\xi)\rho^{\xi-2}\Gamma\right]
+\epsilon^{\xi}\left[\Gamma_{\rho\rho}+(d+1)\frac{1}{\rho}
\Gamma_{\rho}\right]=0.
\end{array}
\end{equation}
with $\rho=r/L(t)$ and $\epsilon\equiv\ell_{\kappa}/L(t).$ In the outer
range where $\rho=O(1)$, the dominant balance in the equation
(\ref{outerscaling}) is between the first term from the time-derivative,
which acts like a forcing, and the second term from the turbulent advection.
The third term may be neglected for small $\epsilon$ , yielding the
leading-order equation for the  outer solution.
This is the same equation which was examined in the preceding section
\ref{similar}, where all self-similar solutions were found. Thus the outer
solutions
$\Gamma_{out}(\rho)$ are given by the formula (\ref{Hcal}) for any choice
of $\alpha$ and multiplied by an arbitrary constant $C_{out}.$ These solutions
must now be matched to an appropriate inner solution in the resistive range.

We introduce the inner variable $\sigma=r/\ell_{\kappa}\equiv \rho/\epsilon$
in Eq. (\ref{outerscaling}) to obtain
\begin{equation}\label{innerscaling}
\begin{array}{lll}
&&\epsilon^{\gamma}\left[\alpha\Gamma+\sigma \Gamma_{\sigma}\right]
+\left[(d-1)\sigma^{\xi}\Gamma_{\sigma\sigma}+
(2\xi+d^2-1)\sigma^{\xi-1} \Gamma_{\sigma}\right.\vspace{2mm}\\
&&\left.+\xi(d-1)(d+\xi)\sigma^{\xi-2}\Gamma\right]
+\left[\Gamma_{\sigma\sigma}+(d+1)\frac{1}{\sigma}\Gamma_{\sigma}\right]=0.
\end{array}
\end{equation}
The dominant balance in (\ref{innerscaling}) is between the second term from
the
turbulent advection and the third term from the molecular resistivity. To
leading order
we can disregard the first term proportional to $\epsilon^{\gamma}$ to get
\begin{equation}
\begin{array}{lll}
\sigma^2\Gamma_{\sigma\sigma}
+(d+1)\sigma\Gamma_{\sigma}
+\sigma^{\xi}
\left[(d-1)\sigma^2 \Gamma_{\sigma\sigma}\right.\vspace{2mm}\\
\left.+(2\xi+d^2-1)\sigma\Gamma_{\sigma}
+\xi(d-1)(d+\xi)\Gamma\right]=0.
\end{array}
\end{equation}
Making the change of variables $y=-(d-1)\sigma^{\xi}$ reduces this to a
hypergeometric equation (\cite{Erdelyi53}, Ch.II):
\begin{equation}
y(1-y)\Gamma_{yy}+\left[c_*-(a_*+b_*+1)y\right]
\Gamma_{y}-a_*b_*\Gamma=0
\end{equation} where
\begin{equation}\label{a+b,ab}
a_*+b_*= \frac{1}{\xi}\left(\frac{2\xi}{d-1}+d\right), \,\,\,\,\,
a_*b_*=c_*=\frac{d+\xi}{\xi}.
\end{equation}
Up to an overall multiplicative constant, there is a unique solution
$F(a_*,b_*;c_*;y)$
of the  above equation which is analytic in the region ${\rm arg}(1-y)<\pi$ of
the complex $y$-plane. This hypergeometric function is given for $|y|<1$
by the absolutely convergent power series,
\begin{equation}
F(a_*,b_*;c_*;y)=\sum_{n=0}^{\infty}
\frac{(a_*)_n(b_*)_n}{(c_*)_n}\frac{y^n}{n!},
\end{equation}
if $c_*\neq0,-1,-2,\ldots$. The other independent solution,
$y^{1-c_*}F(a_*+1-c_*,b_*+1-c_*;2-c_*;y)$ [\cite{Erdelyi53}, 2.9(17)],
is singular at $y=0$ and must be discarded.  Because of the symmetry
$F(a_*,b_*;c_*;y)=F(b_*,a_*;c_*;y)$ we have freedom in choosing $a_*$ and
$b_*.$
Combining the equations in (\ref{a+b,ab}) yields a quadratic equation for $a_*$
$$ a_*^2- \frac{1}{\xi}\left(\frac{2\xi}{d-1}+d\right)a_*+\frac{d+\xi}{\xi}=0$$
and an identical equation for $b_*.$ It is easy to check that the roots are
just
$-\zeta_1/\xi$ and $-\omega_2/\xi,$ where $\zeta_1,\omega_2$ are the scaling
exponents found in section \ref{line}. We choose $a_*=-\zeta_1/\xi$ and
$b_*=-\omega_2/\xi$. Thus, we obtain
$$
\Gamma_{in}(\sigma)= C_{in}
F\left(-\frac{\zeta_1}{\xi},-\frac{\omega_2}{\xi};\frac{\xi+d}{\xi};
-(d-1)\sigma^{\xi}\right). $$
for the inner solution, with an arbitrary constant $C_{in}$. This solution
gives the
complete description in the resistive range, e.g. implying a magnetic energy
spectrum
$E(k)\propto k^{-(1+\xi)}$ for $\ell_\kappa k\gg 1.$

To match this solution to the outer solution, we must find its asymptotic
behavior
for $\sigma\gg 1.$ This is given by $F(a_*,b_*;c_*;y) \sim
\frac{\Gamma(c_*)\Gamma(b_*-a_*)}{\Gamma(b_*)\Gamma(c_*-a_*)} (-y)^{-a_*}$
as ${\rm Re}\,y\rightarrow -\infty,$ for $a_*<b_*,\,\,a_*\neq c_*+p$ with
$p=0,1,2,\ldots$ [see \cite{Erdelyi53}, 2.1.4(17)] to be
$$ \Gamma_{in}(\sigma) \sim C_{in}\frac{\Gamma(c_*)\Gamma(b_*-a_*)}
{\Gamma(b_*)\Gamma(c_*-a_*)} (d-1)^{\zeta_1/\xi}\cdot \sigma^{\zeta_1} $$
for $\sigma\gg 1.$ This is the same power-law as $\Gamma_{out}(\rho)\sim
C_{out}\rho^{\zeta_1}$ for $\rho\ll 1.$ Equating the inner and outer solutions
$\Gamma_{in}(\sigma)=\Gamma_{out}(\rho)$ in the overlap region
$\epsilon\ll\rho\ll 1$
yields the relationship
$$C_{in}=\frac{\Gamma(b_*)\Gamma(c_*-a_*)}
{\Gamma(c_*)\Gamma(b_*-a_*)} (d-1)^{|\zeta_1|/\xi}\cdot \epsilon^{\zeta_1}
\cdot C_{out}.$$
Notice that the first factor is a numerical constant $B(\xi)$ satisfying
$B(0)=d-1$
and $B(\xi_*)=0,$ and varying smoothly between those limits.

Finally, we obtain the magnetic energy from $E(t)=(d/2)\mathpzc{h}^2(t)
\Gamma_{in}(0)=(d/2)\mathpzc{h}^2(t)C_{in}$ which, with $\epsilon=
\ell_\kappa/L(t)$ and $\mathpzc{h}^2(t)=[L(t)]^{-\alpha},$ gives
$$ E(t) = C_1 [L(t)]^{-(\alpha+\zeta_1)} \ell_\kappa^{\zeta_1} $$
for $C_1=(d/2)B(\xi)\cdot C_{out}.$ This differs from the previous heuristic
estimate
only by a constant factor.

\subsection{Magnetic Induction and Dynamo Order Parameter}\lb{order}

The above arguments are reminiscent of our discussion in subsection \ref{line},
where we emphasized the importance of considering the correlations between
line-vectors advected to the {\it same} point, in order to distinguish between
dynamo and non-dynamo regimes. In fact, the two subjects are intimately
related. As we now show, the ``order parameter'' $\mathcal{R}(t)$ that we
considered in (\ref{R0-def}) can be interpreted as the energy of a certain
self-similar decay solution $\mathcal{C}_{(0)}$ corresponding to an initial
condition which is a random, statistically isotropic but spatially uniform
magnetic
field. Such a random magnetic field has a covariance of the form
$$ \mathcal{C}^{ij}_{(0)}(\br,0) = A\delta^{ij} $$
for a positive real number $A.$ A constant correlation such as above would be
invariant  for an advected scalar, but it is not for a magnetic field. There is
a well-known
physical phenomenon of ``shredding'' \cite{BrandenburgSubramanian05} or
``induction'' \cite{Schekochihinetal07} of a constant magnetic field due to the
stretching term $(\bB\bdot\grad)\bu$ in the evolution equation. Thus, an
initially
constant magnetic field will develop very fine-scale structure by turbulent
induction
and may---in principle---act as a seed field for kinematic magnetic dynamo.

The correlation at later times with the above initial condition is provided by
(\ref{C-prop}), which yields
$$ \mathcal{C}^{ij}_{(0)}(\br,t) = A \int d^d\rho \,\,
\bar{F}^{ij}_{kk}(\brho,0|\br,t). $$
For the limit $\kappa\rightarrow 0$ in the KK model, the scaling relation
(\ref{F-scal}) for $\bar{F}$ then implies that
$$  \mathcal{C}^{ij}_{(0)}(\lambda\br,\lambda^\gamma t) =
\mathcal{C}^{ij}_{(0)}(\br,t).
$$
Thus, $\mathcal{C}_{(0)}$ is a self-similar solution of
$\partial_t\mathcal{C}_{(0)}
=\mathcal{M}\mathcal{C}_{(0)}.$ It is clearly the self-similar solution with
parameter
$\alpha=0$ in our general classification. On the other hand, if we take $A=1/d$
then
also
\begin{eqnarray*}
\mathcal{C}^{ij}_{(0)}(\br,t) &= & \frac{1}{d} \int d^d\rho \,\,
F^{ij}_{kk}(\br,t|\brho,0) \cr
      &=& \langle \delta\ell^i(t)\delta\ell^{\prime j}(t)\rangle_\br.
\end{eqnarray*}
The latter expression denotes the correlation of line-vectors which started as
the same
random unit vector at time 0, at any pair of points, which ended up at time $t$
at points
displaced by $\br.$ We should emphasize that this result is valid for any
divergence-free
advecting velocity field and thus applies as well to incompressible fluid
turbulence.
It immediately follows by summing over $i=j$ and setting $r=0$ that
$$ \mathcal{R}(t) = 2 E_{(0)}(t) $$
where $E_{(0)}(t)$ is the energy of the solution $\mathcal{C}_{(0)}.$

Our analysis in the previous section can be applied to describe the behavior of
$\mathcal{C}_{(0)}.$ The formula $C_L(r,t)=\Gamma(r/L(t))$ holds using the
analytic
expression (\ref{Hcal}) for $\Gamma$ with $\alpha=0$ and
$L(t)=(D_1t)^{1/\gamma},$
valid for all $r\gg\ell_\kappa.$  It is more interesting to consider various
asymptotic
behaviors. The ``permanence of large eddies'' implies that
$$ C_L(r,t) \simeq A,\hspace{5mm} r\gg L(t). $$
In the convective range
$$  C_L(r,t) \simeq A \left(\frac{r}{L(t)}\right)^{\zeta_1}, \hspace{5mm}
       \ell_\kappa\ll r\ll L(t). $$
Finally, for $r\rightarrow 0$ and long times,
$$ E_{(0)}(t) \propto A \ell_\kappa^{\zeta_1} (D_1t)^{|\zeta_1|/\gamma} $$
which grows with decreasing $\kappa$ or increasing $t,$ but only as a modest
power law.

This result may be interpreted in terms of material-line correlations
by setting $A=1/d:$
\be \mathcal{R}(t) =\langle\delta\bell(t)\bdot\delta\bell'(t)\rangle_0
      \propto \ell_\kappa^{\zeta_1} (D_1t)^{|\zeta_1|/\gamma},  \lb{R-grow} \ee
which implies that this quantity grows slowly with time. It would be of great
interest to determine the time-dependence also in the dynamo regime.
If the leading eigenfunction of $\mathcal{M}^*$ satisfies $\int d^d r\,
\bar{\mathcal{E}}_{k\ell}(\br)=0,$ then $\mathcal{R}(t)$ need not grow
exponentially.
Note, for example, that the space-integral of the dual eigenfunction
$\mathcal{E}^{ij}(\br)$
does vanish, so the issue is not straightforward.

\section{Final Discussion}

Our work leads to several important conclusions regarding the small-scale
turbulent
kinematic dynamo.

\subsection{Breakdown of Flux Freezing and Dynamo}

In order to understand the turbulent dynamo process a crucial fact is that
magnetic
field lines are {\it not} frozen into the plasma flow, even in the
zero-resistance limit
$\kappa\rightarrow 0.$ Flux-freezing would imply that only a single field line
is
advected into each space point from the field configuration at an earlier time.
In fact, infinitely-many field lines are carried to each point by a combination
of
fluid advection and resistive diffusion \cite{Celanietal06,Eyink09}. In the
Kraichnan
velocity ensemble,  the probability  for two line elements to arrive at the
same point
at time $t$ starting from points separated by $\br$ at time $0$ is
$P(\bzed,t|\br,0)
\propto \exp(-r^\gamma/\gamma D_1t)$ in the limit $\kappa\rightarrow 0$
and does not degenerate into a delta function $\delta^d(\br)$
\cite{Bernardetal97}.
This is a manifestation of the phenomenon of  ``spontaneous stochasticity''
first pointed
out by Bernard et al. \cite{Bernardetal97}, which is due to the explosive
separation
of pairs of fluid particles undergoing turbulent Richardson diffusion. It was
argued
in \cite{Eyink07} that this behavior as $\kappa\rightarrow 0$ holds in general
for a turbulent plasma with a rough velocity field, so that Alfv\'{e}n's
theorem on
flux-conservation remains as a stochastic law only.

The breakdown of flux-freezing in the case of rough velocity fields renders the
turbulent kinematic dynamo an even more subtle problem than the laminar (or
large
Prandtl number) kinematic dynamo (for the latter, see e.g. refs.
\cite{KulsrudAnderson92,
Chertkovetal99,Schekochihinetal02a,Schekochihinetal02b,Schekochihinetal02c}.)
For the very smooth velocities considered there ($\xi=2$), Alfv\'{e}n's theorem
holds
in its usual form in the limit $\kappa\rightarrow 0.$ However, for rougher
velocities
with rugosity exponent anywhere in the range $0<\xi<2,$ an infinite number of
field
lines enters each point even in the zero-resistance limit. The resultant
magnetic
field is the resistive average over the field vectors of all of the individual
lines.
We have shown in this work that the presence of small-scale kinematic dynamo
effect depends upon the existence of sufficient angular correlation between the
individual field vectors. Thus, dynamo action occurs in the KK model for
smoother
velocities with $\xi_*<\xi<2$ but not for rougher velocities with
$0<\xi<\xi_*.$ This
is true despite the fact that the stretching rate of individual fields lines is
much
greater for $\xi$ smaller.

In section \ref{line} we defined
$\mathcal{R}_{k\ell}(\br,t)=F^{ii}_{k\ell}(\bzed,t|\br,0),$
which measures the correlation between line-elements $\delta\bell_k(t)$ and
$\delta\bell_\ell(t)$ at the {\it same} point at time $t$ which started out as
unit
vectors $\hat{{\bf e}}_k,\,\hat{{\bf e}}_\ell$ at {\it distinct} points
separated by $\br$
at time $0.$ We found there that, in the non-dynamo regime of the KK model with
$0<\xi<\xi_*,$  this quantity scales as (\ref{R-scal})
$$ \mathcal{R}_{k\ell}(\br,t) \sim C \ell_\kappa^{\zeta_1}
     (D_1 t)^{-\frac{d+\bar{\zeta}}{\gamma}} \bar{Z}_{k\ell}(\br), $$
for $\ell_\nu,\ell_\kappa\ll r\ll (D_1 t)^{1/\gamma}.$ Here $\bar{Z}$ is an
appropriate
zero-mode of $\mathcal{M}^*$ scaling as $\bar{Z}_{k\ell}(r)\propto
r^{\bar{\zeta}},$
with $-d<\bar{\zeta}<0.$ [Note that the factor $\ell_\kappa^{\zeta_1}$ arises
from
$W^{ii}(\bzed,1)$ in the slow-mode expansion.]  This correlation decays only as
a
power for $r$ increasing through the inertial-convective range, implying that
line vectors initially separated by distances $\sim (D_1 t)^{1/\gamma}$
contribute
substantially to the final average. The correlation does not vanish as
$\kappa\rightarrow
0$ but, in fact, increases as a moderate power of $\ell_\kappa,$ demonstrating
that infinitely-many field lines continue to contribute in that limit.  The
result is, however,
a correlation $\mathcal{R}_{k\ell}(\br,t)$ slowly decaying in time. On the
other hand,
the lengths of the individual line-elements $\langle\delta^2\ell_k(t)\rangle,$
 $\langle\delta^2\ell_\ell(t)\rangle$ grow exponentially in time as in
(\ref{line-grow})
with rate $\lambda\propto \nu/\ell_\nu^2=1/t_\nu.$ The result is that
\be \frac{\mathcal{R}_{k\ell}(\br,t)}
{\sqrt{\langle \delta\ell^2_k(t)\rangle\langle \delta\ell^2_\ell(t)\rangle}}
\rightarrow 0, \lb{decorrelate} \ee
exponentially rapidly either as $t\rightarrow\infty$ or as $\kappa\rightarrow
0$ with
$\nu<Pr_c\kappa.$ We conclude that the dynamo fails for a very rough velocity
field
because advected line-vectors arrive at the same point with insufficient
angular correlation.
Although individual field-lines are stretched to an incredible degree,
resistive averaging
of nearly uncorrelated lines leads to almost complete cancellation.

The situation is qualitatively different in the dynamo range with smoother
velocities ($\xi_*<\xi<2$).  In that case, we have from (\ref{R-exp}) that
$$\mathcal{R}_{k\ell}(\br,t)\propto e^{E_0 t} \bar{\mathcal{E}}_{k\ell}(\br),
$$
where $E_0$ is the dynamo growth rate. Since $E_0\propto 1/t_\kappa\ll
\lambda \propto 1/t_\nu,$ for $\lambda$ in (\ref{Kr74}), it is still true that
the
angular correlations (\ref{decorrelate}) decay exponentially either as
$t\rightarrow\infty$
or as $\kappa\rightarrow 0$ with $Pr$ small enough. However, the decay exponent
is reduced by a finite amount. Enough correlations remain between line-elements
entering a point that the net magnetic field after resistive averaging can
profit
from stretching of individual lines and exponential growth of magnetic energy
ensues.

\subsection{Hydrodynamic and MHD turbulence}

Much of the formalism of this paper carries over to the problem of kinematic
dynamo
for a weak seed magnetic field in hydrodynamic turbulence. The propagators
$\bar{F}^{ij}_{k\ell}(\brho,t|\br,0)=F^{ij}_{k\ell}(\br,0|\brho,t)$ give a
fundamental
description of the kinematic dynamo for any incompressible advecting flow.
All of the results of section \ref{kin-dyn} apply in general, in particular
equations
(\ref{C-prop}), (\ref{F-def}) and (\ref{G-prop}), and also the relationship in
section
\ref{order} between magnetic induction and line-vector correlations. Any
further
simplifications from space-homogeneity and time-stationarity also apply where
appropriate. On the other hand, some features of the KK model are quite special
and do not apply more generally. The self-similarity property (\ref{F-scal}) of
the
propagators $F$ and $\bar{F}$ does not carry over to hydrodynamic turbulence,
because of small-scale intermittency of the advecting velocity field. Also, the
statistics of forward and backward Lagrangian trajectories are not identical
in hydrodynamic turbulence \cite{Sawfordetal05}. Thus, relations such as
(\ref{G-prop2}) which depend upon time-reversal symmetry do not apply to
the real turbulent dynamo. Lastly, the time-evolution of the propagators $F$
and $\bar{F}$ is in general non-Markovian and thus the simple diffusion
equations
such as (\ref{Fbar-eq}),(\ref{F-eq}) do not apply. One of us (G.E.) is
currently carrying
out a numerical evaluation of these propagators for hydrodynamic turbulence,
which
will be reported elsewhere.

We expect that many of the ideas of this work will apply even to nonlinear MHD
turbulence and dynamo effect there. A stochastic form of flux-freezing and
Alfv\'{e}n's theorem holds also for non-ideal (viscous and resistive) MHD
\cite{Eyink09}. We expect these conservation laws to remain stochastic in the
limit
$\kappa\rightarrow 0,\nu\rightarrow 0$ with $Pr$ fixed \cite{Eyink07}. However,
there will be nontrivial differences from the kinematic problem studied here,
due to backreaction of the magnetic field on the plasma flow via the Lorentz
force.
For example, in comparison with hydrodynamic turbulence, 2-particle relative
diffusion
in MHD turbulence is observed to be suppressed in the direction transverse to
the
local magnetic field \cite{Homannetal07}. In principle, however, one can
account for all
such nonlinear effects by the presence of a second stochastic conservation law,
a modified
Kelvin theorem \cite{KuznetsovRuban00,BekensteinOron00,Eyink09}. We believe
that
``spontaneous stochasticity'' and the implied stochasticity of magnetic-line
motion and
flux-conservation must play a central role in the understanding of MHD
turbulence,
dynamo and reconnection.

\vspace{2mm}

\noindent {\bf Acknowledgements.}
We thank W. Hacker for a useful discussion of the matched asymptotics in
section
\ref{energy}.  The work of G. E. was partially supported by NSF grant
AST-0428325
at the Johns Hopkins University. A. F. N. was partially supported by
CNPq-Brazil.

\newpage

\section{Appendix: Slow Mode Expansion for Non-Hermitian Evolution}

Unlike for the passive scalar, the $n$-body evolution operators $\mathcal{M}_n$
for the
passive magnetic field in the Kraichnan model, are no longer even formally
Hermitian,
with $\mathcal{M}_n^*\neq\mathcal{M}_n.$  Nevertheless, certain important
properties
of the scalar evolution operators remain true for $\mathcal{M}_n$ and
$\mathcal{M}_n^*$:
these are homogeneous of degree $-\gamma,$ reality-preserving, and---in the
non-dynamo regime---having absolutely continuous spectrum on the negative real
axis.
As we shall show in the following, the above properties together with assumed
analyticity
conditions allow us to generalize the zero-mode and slow-mode expansions
derived
in \cite{Bernardetal97} for the Hermitian case to pairs of non-Hermitian
operators
$\mathcal{M}$ and $\mathcal{M}^*.$ Although our intended application is to the
Kazantsev-Kraichnan kinematic dynamo model, we shall carry out the derivation
in
an abstract, general setting. We shall employ the properties of $\mathcal{M}$
and $\mathcal{M}^*$
given above and, also, other properties that will be  stated explicitly below.
The entire
argument is modelled very closely after that in \cite{Bernardetal97}, with just
a few important
differences that  are stressed below.

\subsection{The Zero-Mode Expansion}

We shall assume that the operators $\mathcal{M}$ and $\mathcal{M}^*$ act on
$L^2(\mathbb{R}^d).$ The dimension $d$ need not be identified with the physical
space
dimension, as in the main text of the paper. (E.g. if $d_S$ is the space
dimension, then the
$n$-body operators in the Kraichnan model act on $L^2(\mathbb{R}^d)$ with
$d=nd_S$ or
with $d=(n-1)d$ in the translation-invariant  sector.) Define Green's functions
for the
operators by
\begin{equation}
\begin{array}{lll}
 -{\cal M}_{\bx} G(\bx,\by) &=& -{\cal M}_{\by}^*G(\bx,\by)=\delta^d(\bx-\by),
 \,\,\,\,\\
      -{\cal M}_{\bx}^* \bar{G}(\bx,\by) &=& -{\cal M}_{\by}
\bar{G}(\bx,\by)=\delta^d(\bx-\by),
      \end{array}
      \end{equation}
where the subscript ($\bx$ or $\by$) indicates on which variable the operator
acts.
Note that these Green's functions are both real-valued and, of course,
$\bar{G}(\bx,\by)=G(\by,\bx).$

Our aim is to derive the following short-distance asymptotic expansion for $G:$
\begin{equation}
G(\bx/L,\by) \sim \sum_a L^{-\zeta_a}
f_a(\bx)[\bar{g}_a(\by)]^*,
      \,\,\,\,\, L\gg 1,
      \lb{eq5}
      \end{equation}
where $^{*}$ here denotes complex-conjugation. The function $f_a$
is a {\it regular zero-mode} of ${\cal M}$ with scaling dimension
$\zeta_a,$ while $\bar{g}_a$ is a {\it singular zero-mode} of
${\cal M}^*$ with scaling dimension
$\bar{\omega}_a=-d+\gamma-\zeta_a^*.$ What dominates in the
expansion (\ref{eq5}) is the contributing zero-mode whose scaling exponent
$\zeta_a$ has the smallest real part. Thus, we label the exponents according
to the magnitude of their real part, so that ${\rm Re}\,\zeta_a>{\rm
Re}\,\zeta_{a'}$
and ${\rm Re}\,\omega_a<{\rm Re}\,\omega_{a'}$ for $a>a'$.
We derive also a similar expansion
for the adjoint Green's function
\begin{equation}
\bar{G}(\bx/L,\by) \sim \sum_a L^{-\bar{\zeta}_a}
\bar{f}_a(\bx)[g_a(\by)]^*,
      \,\,\,\,\, L\gg 1,
      \lb{eq6} \end{equation}
where now  $\bar{f}_a$ is a {\it regular zero-mode} of ${\cal M}^*$
with scaling dimension $\bar{\zeta}_a,$ while $g_a$ is a {\it
singular zero-mode} of ${\cal M}$ with scaling dimension
$\omega_a=-d+\gamma-\bar{\zeta}_a^*.$  We thus see that the
homogeneous zero-modes of the operators ${\cal M}$ and ${\cal
M}^*$ come in pairs, $(\bar{f}_a, g_a)$ and $(f_a, \bar{g}_a)$, with
related scaling exponents.


Following \cite{Bernardetal97}, we employ the Mellin transform, which
is a unitary transformation between the spaces
$L^2(\mathbb{R}^{d})$ and $L^{2}(Re\,\sigma=-d/2)\otimes
L^{2}(S^{d-1})$ given by
\begin{equation}
f(\bx)\mapsto \tilde{f}(\sigma,\hat{\bx})
=\int_0^{\infty}\lambda^{-\sigma-1}f(\lambda \hat{\bx})d\lambda.
\end{equation}
with the inverse transform, for $R=|\bx|,$
\begin{equation}\label{Finverse}
f(\bx)=\frac{1}{2\pi
i}\int_{Re\,\sigma=-\frac{d}{2}}R^{\sigma}\tilde
{f}(\sigma,\hat{\bx})d\sigma.
\end{equation}
The inner product on $L^{2}(Re\,\sigma=-d/2)\otimes
L^{2}(S^{d-1})$ is:
\be\label{IP}
\left<\tilde{f},\tilde{g}\right>=\frac{1}{2\pi i} \int d\omega(\hat{\bx})
\int_{Re\, \sigma=-\frac{d}{2}}d\sigma
\left[\tilde{f}(\sigma,\hat{\bx})\right]^*
\tilde{g}(\sigma,\hat{\bx})
\ee
However, it is more convenient to write this as
\be\lb{IP*}
\left<\tilde{f},\tilde{g}\right>=\frac{1}{2\pi i} \int d\omega(\hat{\bx})
\int_{Re\,
\sigma=-\frac{d}{2}}d\sigma
\left[\tilde{f}(-\sigma^*-d,\hat{\bx})\right]^*
\tilde{g}(\sigma,\hat{\bx}).
\ee
Although $\left[\tilde{f}(\sigma,\hat{\bx})\right]^*=
\left[\tilde{f}(-\sigma^*-d,\hat{\bx})\right]^*$ on the line ${\rm Re}\,\sigma=
-d/2,$ the second expression is analytic in $\sigma$ when $\tilde{f}(\sigma,
\hat{\bx})$ is analytic. This form of the inner product allows one to shift
integration
contours in the complex $\sigma$-plane.

A key role in the analysis is played by the operator
\begin{equation}\label{N}
{\cal N}=R^{\gamma/2}{\cal M}R^{\gamma/2}
\end{equation}
which is homogeneous of degree zero. Since it thus commutes with the
self-adjoint
generator $D=\frac{1}{i}\left(\bx\bdot\grad_\bx+\frac{d}{2}\right)$ of
dilatations,
it is partially diagonalized under the Mellin transform:
$$ (\mathcal{N}f)^\sim(\sigma,\hat{\bx})=\tilde{\mathcal{N}}(\sigma)
      \tilde{f}(\sigma,\hat{\bx}), $$
where $\tilde{{\mathcal{N}}}(\sigma)$ for each $\sigma$ is an operator on
$L^{2}(S^{d-1}).$   Using ${\cal M}^{-1}=R^{\gamma/2}{\cal
N}^{-1}R^{\gamma/2},$ one
straightforwardly derives the following fundamental identity for the Green's
function
$G(\bx,\by)=-{\cal M}^{-1}(\bx,\by)$:
\begin{eqnarray}
G(\bx,\by)&=&\displaystyle-\int_{Re \sigma=-\frac{d}{2}+\frac{\gamma}{2}}
\frac{d\sigma}{2\pi i} \left[R(\bx)\right]^{\sigma}
\tilde{\cal
N}^{-1}\left(\sigma-\frac{\gamma}{2};\hat{\bx},\hat{\by}\right)\vspace{2mm}\cr
&&\hspace{3cm}\times\left[R(\by)\right]^{-d+\gamma-\sigma}.
\lb{fund} \end{eqnarray}
See \cite{Bernardetal97}. We note that the shifts in $\sigma$ arise because
$R^{\gamma/2}$ acts as a translation by $-\gamma/2$ under the
Mellin transform. The above identity is the key to deriving the zero-mode
expansion for $G$.

The main hypothesis is that the operator function $\tilde{\cal N}^{-1}(\sigma)$
is meromorphic in a wide vertical strip around the line ${\rm
Re}\,\sigma=-d/2,$
whose only singularities are poles
$$
-\widetilde{\cal
N}^{-1}\left(\sigma-\frac{\gamma}{2},\hat{\bx},\hat{\by}\right)
\cong
\frac{Z_a(\hat{\bx},\hat{\by})}{\sigma-\zeta_a}
$$
at complex values $\zeta_a,\,\,\,a=1,2,\ldots$ in the strip. By moving
the integration contour in (\ref{fund}) further and further to the right, one
picks up successive pole contributions. This implies that Green's function
for large $L$ satisfies:
$$
G\left(\frac{\bx}{L},\by\right)=\sum_a L^{-\zeta_a} Z_a(\bx,\by) $$
with the function
$$
Z_a(\bx,\by)\equiv \left[R(\bx)\right]^{\zeta_a} Z_a(\hat{\bx},\hat{\by})
[R(\by)]^{-d+\gamma-\zeta_a}
$$
which is homogeneous of degree $\zeta_a$ in $\bx$ and of degree
$-d+\gamma-{\zeta}_a$ in $\by$. From the definition of the Green's function,
using ${\cal M}_{\bx}=L^{-\gamma}{\cal M}_{\bx'}$ with $\bx'=\bx/L$,
$$\begin{array}{lll}
-L^{-\gamma}\delta^{d}\left(\frac{\bx}{L}-\by\right)&=&{\cal
M}_{\bx}
G\left(\frac{\bx}{L},\by\right)\vspace{2mm}\\
&=&\sum_aL^{-\zeta_a}{\cal
M}_{\bx} Z_a(\bx,\by)
\end{array}$$
from which we get ${\cal M}_{\bx}Z_a(\bx,\by)=0$ for points off the diagonal.
Likewise,
$$\begin{array}{lll}
-\delta^{d}\left(\frac{\bx}{L}-\by\right)&=&{\cal M}_{\by}^*
G\left(\frac{\bx}{L},\by\right)\vspace{2mm}\\
&=&\sum_aL^{-\zeta_a} \mathcal{M}_{\by}^{*}Z_a(\bx,\by)
\end{array}$$
from which we get ${\cal M}_{\by}^*Z_a(\bx,\by)=0$ for points off the diagonal.
We finally conclude that $Z_a(\bx,\by)$ for fixed $\by$ is a homogeneous
zero-mode
of $\mathcal{M}_\bx$ of degree $\zeta_a$ and for fixed $\bx$ is a homogeneous
zero-mode of $\mathcal{M}_\by^*$ of degree $-d+\gamma-{\zeta}_a.$  If we assume
that zero-modes of a given scaling exponent are non-degenerate, as will
generically
be true, then we can write
$$ Z_a(\bx,\by)=f_a(\bx) \left[\bar{g}_a(\by)\right]^*, $$
where $f_a$ is the unique scaling zero-mode of $\mathcal{M}$ with exponent
$\zeta_a$
and $\bar{g}_a$ is the scaling zero-mode of $\mathcal{M}^*$ with exponent
$\bar{\omega}_a=-d+\gamma-{\zeta}_a^*.$ We have used here the fact that
$\mathcal{M}^*$
is reality-preserving. This yields (\ref{eq5}). The expansion (\ref{eq6}) for
$\bar{G}$ is
derived by an identical argument.

Although we shall not employ the corresponding large-distance expansion
in this work, we make here a few remarks about it. Under the Mellin transform
the adjoint of $\mathcal{N}^{-1}$ has the kernel
$$
\widetilde{{\cal
N}^*}^{-1}(\sigma;\hat{\bx},\hat{\by})=\left[\widetilde{\cal N}^{-1}
(-\sigma^*-d;\hat{\by},\hat{\bx})\right]^*.
$$
This last relation reveals the important fact that if $\tilde{{\cal
N}}^{-1}(\sigma)$ has
a pole at $\zeta_a$ then $\widetilde{{\cal N}^*}^{-1}(\sigma)$ has a pole at
$\bar{\omega}_a=-d+\gamma-\zeta_a^*$. Indeed we have:
$$
\begin{array}{lll}
\widetilde{{\cal N}^*}^{-1}\left(\sigma-\frac{\gamma}{2};
\hat{\bx},\hat{\by}\right)&=&\left[\tilde{\cal N}^{-1}\left(-d+\frac{\gamma}{2}
-\sigma^*;\hat{\by},\hat{\bx}\right)\right]^*\vspace{2mm}\\
&=&\left[\tilde{\cal N}^{-1}(-d+\gamma-\sigma^*-\frac{\gamma}{2};
\hat{\by},\hat{\bx})\right]^*\vspace{2mm}\\
&\cong&\left\{\frac{-1}{(-d+\gamma-\sigma^*)-\zeta_a}
f_a(\hat{\by})\left[\overline{g}_a({\hat{\bx}})\right]^*
\right\}^*\vspace{2mm}\\
&\cong&\frac{-1}{\bar{\omega}_a-\sigma}
\overline{g}_a({\hat{\bx}})\left[f_a(\hat{\by})\right]^*.
\end{array}$$
At this pole with ${\rm Re}\,\bar{\omega}_a<-d/2$ the role of the regular and
singular
zero-modes in the residue is reversed. By pushing the integration contour in
(\ref{fund})
further and further to the left one can thus derive a large-distance expansion
for $\bar{G}$.
This can also be directly obtained from the short-distance expansion for $G$,
as follows:
$$
\begin{array}{lll}
\bar{G}(L\bx,\by)&=&\bar{G}\left(L\bx,L\frac{\by}{L}\right)
\vspace{2mm}\\
&=&L^{\gamma-d}\bar{G}\left(\bx,\frac{\by}{L}\right)
\vspace{2mm}\\
&=&L^{\gamma-d}\left[G\left(\frac{\by}{L},\bx\right)\right]^*
\vspace{2mm}\\
&=&L^{\gamma-d}\displaystyle\sum_a L^{-\zeta^*_a}
\left[f_a(\by)\right]^*
\bar{g}_a(\bx)\vspace{2mm}\\
&=&\displaystyle\sum_a L^{\bar{\omega}_a}\bar{g}_a(\bx)\left[f_a(\by)\right]^*
\end{array}
$$
for $L\gg1$. Of course, a similar expansion holds for $G.$

\subsection{The Slow Mode Expansion:  Elementary Arguments}

Define the heat-kernels
\begin{equation}
\begin{array}{lll} P(\bx,t|\bx_0,t_0)&=&\langle\bx|e^{(t-t_0){\cal
M}}|\bx_0\rangle,
\,\,\,\,\\
\bar{P}(\bx,t|\bx_0,t_0)&=&\langle\bx|e^{(t-t_0){\cal
M^*}}|\bx_0\rangle,
 \end{array}
 \end{equation} so that, obviously,
$\bar{P}(\bx,t|\bx_0,t_0)=P(\bx_0,t|\bx,t_0).$
We have the relations
\be
G(\bx,\by)=\int_0^\infty dt\, P(\bx,t|\by,0)
\lb{G-P} \ee
and
\be
\bar{G}(\bx,\by) =\int_0^\infty dt\, \bar{P}(\bx,t|\by,0).
\lb{Gbar-Pbar} \ee
Given the validity of the zero-mode expansions for $G$ and $\bar{G}$ one
should expect that related expansions hold for $P$ and $\bar{P}.$
We shall show that this is indeed true, with the asymptotic expansion
analogous to (\ref{eq5}) for $L\gg 1:$
\begin{equation}
P\left(\frac{\bx}{L},t\big|\bx_0,0\right)=  \sum_{a, p\geq 0}
L^{-(\zeta_a+\gamma p)}
      f_{a,p}(\bx)[\bar{g}_{a,p}(\bx_0,t)]^*.
      \lb{eq8} \end{equation}
Here $f_{a,p}$ are the {\it tower of regular slow modes} of ${\cal
M},$ satisfying $-{\cal M}f_{a,p}=f_{a,p-1}$ and $f_{a,0}=f_a.$
Also, $\bar{g}_{a,p}$ are solutions of
$\partial_t\bar{g}_{a,p}(\bx,t)={\cal M}^*\bar{g}_{a,p}(\bx,t)$ with
initial conditions $\bar{g}_{a,-1}(\bx,0)=\bar{g}_a(\bx)$ and
$\bar{g}_{a,p+1}(\bx,0)=-{\cal M}^* \bar{g}_{a,p}(\bx,0).$ They
satisfy the scaling relations
$\bar{g}_{a,p}(\lambda\bx,\lambda^\gamma
t)=\lambda^{\bar{\omega}_a-(p+1)\gamma}\bar{g}_{a,p}(\bx,t).$ Note that
the dominant contribution in (\ref{eq8}) will generally come from the
tower with minimum ${\rm Re}(\zeta_a)$ and from the first
(zero-mode) term $p=0.$ There is an analogous expansion for
$\bar{P}$ with $L\gg 1$:
\begin{equation}
 \bar{P}\left(\frac{\bx}{L},t\big|\bx_0,0\right)=
 \sum_{a, p\geq 0} L^{-(\bar{\zeta}_a +\gamma p)}
      \bar{f}_{a,p}(\bx)[g_{a,p}(\bx_0,t)]^*,
    \lb{eq9} \end{equation}
with the roles of the operators ${\cal M}$ and ${\cal M}^*$ reversed.

We shall derive the above expansions in this section and the next.
Here we proceed by assuming that a general expansion exists
for $L\gg 1$ of the form
\be
P\left(\frac{\bx}{L},t\,\vrule\,\bx',0\right) \cong \sum_\alpha
L^{-\rho_\alpha} f_\alpha(\bx)\left[\bar{g}_\alpha(\bx',t)\right]^*.
\lb{assume} \ee
We shall then identify the form this expansion must take. In the following
section we establish from a more fundamental point of view the existence
of such an expansion.

First we substitute (\ref{assume}) into
$$\begin{array}{lll}
\partial_tP\left(\bx,t|\bx',0\right)&=&{\cal
M}_{\bx'}^*P\left(\bx,t|\bx',0\right)
\vspace{2mm}\\
&=&{\cal M}_{\bx}P\left(\bx,t|\bx',0\right),
\end{array}$$
obtaining
\begin{eqnarray}
&&\sum_\alpha L^{-\rho_\alpha}f_\alpha(\bx)
\left[\partial_t\bar{g}_\alpha(\bx',t)\right]^* \vspace{2mm}\cr
&&=\sum_\alpha L^{-\rho_\alpha}f_\alpha(\bx)
\left[{\cal M}_{\bx'}^*\bar{g}_\alpha(\bx',t)\right]^*\vspace{2mm}\cr
&&=\sum_\alpha L^{-\rho_\alpha+\gamma}{\cal M}_{\bx}f_\alpha(\bx)
\left[\bar{g}_\alpha(\bx',t)\right]^*.
\lb{ev-cond} \end{eqnarray}
We see that whenever the asymptotic series contains a term proportional
to $f_\alpha(\bx)$ with scaling exponent $\rho_\alpha$ it must also contain
a term ${\cal M}_{\bx}f_\alpha(\bx)$ with exponent $\rho_\alpha-\gamma,$
and then a term ${\cal M}_{\bx}^2f_\alpha(\bx)$ with exponent
$\rho_\alpha-2\gamma,$ and so on. This cannot continue indefinitely,
since, otherwise, there would be successively more and more divergent terms
for $L\gg 1.$ The only way that this sequence can terminate is if, eventually,
$$ {\cal M}_{\bx}^{p+1} f_\alpha(\bx)=0 $$
for some integer $p.$ In that case, we see that $f_\alpha=(-{\cal
M}_{\bx})^pf_a
\equiv f_{a,p}$ for some homogeneous zero-mode $f_a,$ and the expansion
(\ref{assume}) contains the whole tower above that zero mode.  All such
towers  associated to regular zero modes must appear because the condition
(\ref{G-P}) together with the zero-mode expansion for $G$ implies that
\begin{eqnarray}
&& \sum_\alpha L^{-\rho_\alpha}f_\alpha(\bx)
\left[\int_0^\infty dt\,\bar{g}_\alpha(\bx',t)\right]^* \cr
&& \hspace{1cm} \cong
\sum_a L^{-\zeta_a} f_a(\bx) \left[\bar{g}_a(\bx')\right]^*.
\lb{G-cond} \end{eqnarray}
The expansion (\ref{assume}) thus must have precisely the form of
equation (\ref{eq8}) and we must only establish the properties of
$\bar{g}_\alpha=\bar{g}_{a,p}.$ We note from (\ref{ev-cond}) that
$$ \partial_t\bar{g}_{a,p}=\mathcal{M}^*\bar{g}_{a,p}=-\bar{g}_{a,p+1}. $$
Also (\ref{G-cond}) implies that (away from the origin $\bx'=\bzed$)
\begin{eqnarray*}
\bar{g}_{a,p-1}(\bx',0) &=& -\int_0^\infty dt\, \partial_t
\bar{g}_{a,p-1}(\bx',t) \cr
                           &=&  \int_0^\infty dt\, \bar{g}_{a,p}(\bx',t) =0
\end{eqnarray*}
for $p=1,2,3\ldots,$ whereas $\bar{g}_{a,-1}(\bx',0)=\bar{g}_a(\bx'),$
the singular zero-mode of $\mathcal{M}^*.$ Finally, the scaling properties
of $\bar{g}_{a,p}$ follow from the scaling property of $\bar{g}_a$ and of
the propagator, i.e., $\bar{g}_a(\lambda\bx)=\lambda^{\bar{\omega}_a}
\bar{g}_a(\bx)$ and $e^{\lambda^{\gamma}t\cal
M^*}\left(\lambda\bx,\lambda\by\right)
=\lambda^{-d}e^{t{\cal M}^*}(\bx,\by)$, respectively. The expansion (\ref{eq9})
for
$\bar{P}$ is derived by an identical argument.

\subsection{\label{sec:level4} The Slow Mode Expansion: Fundamental
Derivation}

We shall now demonstrate the existence of the expansion (\ref{assume})
and verify by an independent argument its general properties discussed
above. A key fact that we use is that the operators ${\cal M}$ and ${\cal M}^*$
both have spectrum absolutely continuous over the negative real axis.
This assumption explicitly rules out kinematic dynamo effect due
to point spectrum on the positive real axis. Because of this assumed property,
we may define
\begin{equation}
X=\log(-{\cal M}),\,\,\,\,X^*=\log(-{\cal M}^*)
\end{equation} where the branch of the natural logarithm
$\log(z)$ is defined with cut along the negative real axis.
Furthermore, because $\mathcal{M}$ and $\mathcal{M}$ are homogeneous
of degree $-\gamma,$ the operators $X$ and $X^*$ both satisfy
the Heisenberg commutation relations
\begin{equation}
 [D,X]=[D,X^*]=i\gamma I,
\lb{eq2}
\end{equation}
where $D$ is the self-adjoint generator of dilatations. We may decompose
$X,X^*$ into Hermitian and skew-Hermitian parts, as
\begin{equation}
 X=H+iK, \,\,\,\, X^*=H-iK,
\lb{eq3}
\end{equation}
where $H,K$ are both Hermitian. In that case, we see that
\begin{equation}\label{comut} [D,H]=i\gamma
I,\,\,\,\, [D,K]=0.
\end{equation}

We can now follow the arguments in Ref. \cite{Bernardetal97} to infer
that under the unitary Mellin transform \begin{equation}
D\longrightarrow {{1}\over{i}}\left(\sigma+{{d}\over{2}}\right),
\end{equation}
\begin{equation}\label{U} H\longrightarrow \wh{U}(\sigma)
\gamma\partial_\sigma \wh{U}^{-1}(\sigma),
\end{equation}
\begin{equation}\label{K0} K\longrightarrow \wh{K}_0(\sigma)=
\wh{U}(\sigma) \wh{K}(\sigma)
     \wh{U}^{-1}(\sigma),
   \end{equation}
where $Re(\sigma)=-{{d}\over{2}}.$  As in Ref. \cite{Bernardetal97}, the
operators $\wh{U}(\sigma)$ in Eq. (\ref{U}) are unitary operators
on $L^2(S^{d-1})$ and the result in Eq. (\ref{U}) follows from the
Stone-von Neumann theorem on uniqueness of representations of the
Heisenberg algebra. The operators $\wh{K}_0(\sigma)$ in (\ref{K0})
are self-adjoint operators on $L^2(S^{d-1})$ and the result
(\ref{K0}) is a consequence of the second half of
(\ref{comut})---commutativity of $D$ and $K$---so that $K$ leaves
invariant the eigenspaces of $D.$ It is convenient to introduce
instead the self-adjoint operators $\wh{K}(\sigma) =
\wh{U}^{-1}(\sigma) \wh{K}_0(\sigma)\wh{U}(\sigma).$ Thus,
\begin{equation}\label{X} X\longrightarrow
\wh{U}(\sigma)[\gamma\partial_\sigma
      +i\wh{K}(\sigma)]\wh{U}^{-1}(\sigma),
     \end{equation}
\begin{equation}\label{X*} X^*\longrightarrow
\wh{U}(\sigma)[\gamma\partial_\sigma
      -i\wh{K}(\sigma)]\wh{U}^{-1}(\sigma).
      \end{equation}

We now introduce the operators $\wh{L}(\sigma)$ on $L^2(S^{d-1})$
satisfying \begin{equation} \gamma {{d}\over{d\sigma}}
\wh{L}(\sigma) =-i\wh{K}(\sigma)\wh{L}(\sigma),
       \,\,\,\, \wh{L}(0)=I,
     \end{equation}
\begin{equation} \gamma {{d}\over{d\sigma}} \wh{L}^{-1}(\sigma)
=\wh{L}^{-1}(\sigma) i\wh{K}(\sigma),
       \,\,\,\, \wh{L}^{-1}(0)=I.
      \end{equation}
The operators $\wh{L}(\sigma)$ and $\wh{L}^{-1}(\sigma)$ can be
defined explicitly by ordered exponentials along the line
$\sigma=-{{d}\over{2}}+i\nu:$ \begin{equation} \wh{L}(\sigma) =
\left\{\begin{array}{ll}  {\rm Texp}\left[
{{1}\over{\gamma}}\int_0^\nu d\nu'
      \, \wh{K}\left(-{{d}\over{2}}+i\nu'\right)\right]
                     &{\rm if}\,\, \nu \geq 0, \\
     \overline{{\rm T}}{\rm exp}\left[ -{{1}\over{\gamma}}\int_\nu^0 d\nu'
     \, \wh{K}\left(-{{d}\over{2}}+i\nu'\right)\right]
                     &{\rm if}\,\, \nu <0.
                     \end{array}\right.
                     \end{equation}
and \begin{equation}
 \wh{L}^{-1}(\sigma) =
\left\{\begin{array}{ll}
       \overline{{\rm T}}{\rm exp}\left[ -{{1}\over{\gamma}}\int_0^\nu d\nu'
      \, \wh{K}\left(-{{d}\over{2}}+i\nu'\right)\right]
                     &{\rm if}\,\, \nu \geq 0, \\
     {\rm Texp}\left[ {{1}\over{\gamma}}\int_\nu^0 d\nu'
     \, \wh{K}\left(-{{d}\over{2}}+i\nu'\right)\right]
                     &{\rm if}\,\, \nu <0.
                     \end{array}\right.
                     \end{equation} It follows that
\begin{equation}\label{partial+iK}\gamma\partial_\sigma +i\wh{K}(\sigma) =
                   \wh{L}(\sigma) \gamma\partial_\sigma \wh{L}^{-1}(\sigma),
                   \end{equation}
\begin{equation}\label{partial-iK}
 \gamma\partial_\sigma -i\wh{K}(\sigma) =
                   \wh{L}^{* -1}(\sigma) \gamma\partial_\sigma
\wh{L}^{*}(\sigma).
\end{equation}

Finally, combining (\ref{partial+iK}), (\ref{partial-iK}) with
(\ref{X}), (\ref{X*}), we obtain the mappings under the Mellin
transform
\begin{equation}
 X\longrightarrow  \wh{V}(\sigma)\gamma\partial_\sigma \wh{V}^{-1}(\sigma),
\lb{eq16}
 \end{equation}
\begin{equation} X^*\longrightarrow  \wh{V}^{*
-1}(\sigma)\gamma\partial_\sigma \wh{V}^*(\sigma).
\lb{eq17}
\end{equation} with
\begin{equation}
\wh{V}(\sigma)=\wh{U}(\sigma)\wh{L}(\sigma),\,\,\,\,
      \wh{V}^*(\sigma)=\wh{L}^*(\sigma)\wh{U}^{-1}(\sigma).
\end{equation} This is the main result that we require.

The rest of the derivation of the slow mode expansion follows the argument
of Ref.\cite{Bernardetal97}, assuming that $\wh{V}(\sigma)$ extends to a
meromorphic operator-valued function of $\sigma.$ We shall sketch here
the main points. Note first that we can exponentiate the relations
(\ref{eq16}),(\ref{eq17}) to obtain
\be -\mathcal{M}=VR^{-\gamma}V^{-1},\,\,\,\,\,-\mathcal{M}^*=
V^{*\,-1}R^{-\gamma}V^*,
\lb{M-V} \ee
where we have defined the operators $V$ and $V^*$ by
$$(Vf)^\sim(\sigma,\hat{\bx})\equiv \wh{V}(\sigma)f(\sigma,\hat{\bx}), \,\,
      (V^*f)^\sim(\sigma,\hat{\bx})\equiv \wh{V}^*(\sigma)f(\sigma,\hat{\bx}),
$$
which are mutual adjoints. From the definition $ \mathcal{N}^{-1}=
R^{-\gamma/2}\mathcal{M}^{-1}R^{-\gamma/2}$ and (\ref{M-V})
we see that
$$ -\mathcal{N}^{-1}
    =(R^{-\gamma/2}VR^{\gamma/2})(R^{\gamma/2}V^{-1}R^{-\gamma/2}),$$
which under  Mellin transform becomes
\be -\wh{\mathcal{N}}^{-1}\left(\sigma-\frac{\gamma}{2}\right)
      =\wh{V}(\sigma)\wh{V}^{-1}(\sigma-\gamma). \lb{Nt-Vt} \ee
Of course, we have also
 \be -\wh{\mathcal{N}}^{*\,-1}\left(\sigma-\frac{\gamma}{2}\right)
      =\wh{V}^{*\,-1}(\sigma)\wh{V}^{*}(\sigma-\gamma), \lb{Nst-Vst} \ee
by an identical argument.

One immediate consequence of (\ref{Nt-Vt}) is that poles of
$\wh{\mathcal{N}}^{-1}
\left(\sigma-\frac{\gamma}{2}\right)$ can arise only from poles of
$\wh{V}(\sigma)$
or zeroes of $\wh{V}(\sigma-\gamma).$ Our main assumption will be that all of
the
poles of $\wh{V}(\sigma)$ lie in the half-plane ${\rm Re}\,\sigma>-d/2$ and all
of
its zeroes lie in the half-plane ${\rm Re}\,\sigma<-d/2.$ Because of the
adjoint
relation
\be \wh{V}^{*\,-1}(\sigma)=\left[\wh{V}(-\sigma^*-d)\right]^{*\,-1}, \lb{V-adj}
\ee
we see that $\wh{V}^{*\,-1}(\sigma)$ then enjoys the same property, with the
poles
of $\wh{V}(\sigma)$ corresponding to zeroes of $\wh{V}^{*\,-1}(\sigma)$ and
the zeroes of $\wh{V}(\sigma)$ corresponding to poles of
$\wh{V}^{*\,-1}(\sigma).$
The assumption on $\wh{V}$ implies that all of the poles of
$\wh{\mathcal{N}}^{-1}
\left(\sigma-\frac{\gamma}{2}\right)$ for ${\rm Re}\,\sigma>-d/2+\gamma/2$
must arise from poles of $\wh{V}(\sigma)$ with the form
\be \wh{V}(\sigma) \cong \frac{1}{\sigma-\zeta_a}
|f_a\rangle\langle \bar{g}_a|\wh{V}(\sigma_a-\gamma), \lb{V-pole} \ee
in order to reproduce the known poles of $\wh{\mathcal{N}}^{-1}
\left(\sigma-\frac{\gamma}{2}\right).$ On the other hand, we can rewrite
(\ref{Nt-Vt}) as
$$ \wh{V}(\sigma+\gamma p)=
-\wh{\mathcal{N}}^{-1}\left(\sigma+\gamma\left(p-\frac{1}{2}\right)\right)
\wh{V}(\sigma+\gamma(p-1)), $$
for $p=1,2,\ldots.$ Let us assume that none of the poles of
$\wh{\mathcal{N}}^{-1}
\left(\sigma-\frac{\gamma}{2}\right)$ occur at points in the complex
$\sigma$-plane
with real parts differing by integer multiples of $\gamma.$ This will hold
generically.
In that case,
$\wh{\mathcal{N}}^{-1}\left(\sigma_a+\gamma\left(p-\frac{1}{2}\right)\right)$
is a regular operator for all $p=1,2,\ldots$ and we may use the above relation
to
infer inductively a series of poles
$$ \wh{V}(\sigma) \cong \frac{1}{\sigma-\zeta_a-\gamma p}
|f_{a,p}\rangle\langle \bar{g}_a|\wh{V}(\sigma_a-\gamma), $$
for each $a=1,2,\ldots$ with
$$ f_{a,p}=-\wh{\mathcal{N}}^{-1}\left(\sigma_a+
\gamma\left(p-\frac{1}{2}\right)\right)
     f_{a,p-1}$$
for $p=1,2,\ldots$ and $f_{a,0}=f_a.$ It is not difficult to check that this
coincides with
the definition of $f_{a,p}$ given earlier.

Finally, we exponentiate one more time relations (\ref{M-V}) to obtain
\be e^{t\mathcal{M}}=Ve^{-t R^{-\gamma}}V^{-1},\,\,\,\,\,
       e^{t\mathcal{M}^*}=V^{*\,-1}e^{-t R^{-\gamma}}V^*.
\lb{P-V} \ee
The first of these, under Mellin transform, gives
\begin{eqnarray*}
&& (e^{t\mathcal{M}}\varphi)(\hat{\bx}/L)=\cr
&& \hspace{1cm} \frac{1}{\gamma}
\int_{{\rm Re}\,\sigma=-d/2}\frac{d\sigma}{2\pi i} L^{-\sigma}\int
d\omega(\hat{\by})
      \wh{V}(\sigma;\hat{\bx},\hat{\by}) \cr
&& \hspace{1cm} \times \int_{{\rm Re}\,\sigma'=-d/2-0}\frac{d\sigma'}{2\pi i}
       t^{(\sigma'-\sigma)/\gamma}
\Gamma\left(\frac{\sigma-\sigma'}{\gamma}\right)  \cr
&& \hspace{3cm}
       (\wh{V}^{-1}(\sigma')\wh{\varphi})(\sigma',\hat{\by}).
\end{eqnarray*}
Pushing the $\sigma$-integration contour further and further to the right gives
the expansion for $L\gg 1:$
$$ (e^{t\mathcal{M}}\varphi)(\hat{\bx}/L) \cong \sum_{a,p} L^{-\sigma-\gamma p}
      f_{a,p}(\bx) \langle \bar{g}_{a,p}(t),\varphi\rangle, $$
with a suitable definition of $\bar{g}_{a,p}(t)$ See \cite{Bernardetal97} for
more
details. The above is just an integrated form of the slow-mode expansion
(\ref{eq8})
for $P$. The slow-mode expansion (\ref{eq9}) for $\bar{P}$ follows by an
identical argument, in which the various terms arise from the poles of
$\wh{V}^{*\,-1}(\sigma)$ in the half-plane ${\rm Re}\,\sigma>-d/2.$

There are similar large-distance expansions for $P$ and $\bar{P},$ in which
enter the ``tunnels'' of singular slow modes.  The terms in these expansions
arise from the zeroes of $\wh{V}(\sigma')$ and $\wh{V}^{*\,-1}(\sigma')$ in the
half-plane ${\rm Re}\,\sigma'<-d/2$ by moving $\sigma'$-integration contours
to the left in a formula similar to the above. The reader may work out details.
We shall just note here that the pole (\ref{V-pole}) of $\wh{V}(\sigma)$
implies
via the relation
$$ \wh{V}^*(\sigma) =\left[V(-\sigma^*-d)\right]^*$$
the result
\be \wh{V}^*(\sigma-\gamma) \cong \frac{1}{\bar{\omega}_a-\sigma}
\wh{V}^*(\bar{\omega}_a)|\bar{g}_a\rangle\langle f_a| \ \lb{Vs-pole} \ee
and thus the zero of $\wh{V}^{*\,-1}(\sigma)$ at
$\sigma=\bar{\omega}_a-\gamma.$
This zero and the ``tunnel'' of zeroes beneath it give rise to the terms in the
large-distance expansion of $\bar{P}.$

In our discussion throughout the appendix, we have assumed that all the regular
zero
modes of $\mathcal{M}$ and $\mathcal{M}^*$ have scaling exponents $\sigma$ with
${\rm Re}\,\sigma>-d/2+\gamma/2$ and all the singular zero modes have scaling
exponents
with ${\rm Re}\,\sigma<-d/2-\gamma/2.$ This is true in the KK model only for
$d\geq 6$
and for $\xi$ not too large. For $d\leq 4$ the two primary ``singular modes''
have
exponents $\omega_1,\bar{\omega}_1\geq -d/2$ for all $\xi$ and, for
sufficiently large
$\xi,$  these exponents even cross and become larger than
$\zeta_1,\bar{\zeta}_1,$
respectively!  For $d\geq 6$ it still happens that $\zeta_1<-d/2$ and
$\bar{\omega}_1
>-d/2$ for sufficiently large $\xi.$ These results are not consistent with the
assumptions
made in the derivation sketched above. Nevertheless, the zero- and slow-mode
expansions
seem to hold for all $d>2$ and $0<\xi<\xi_*.$

\newpage


%
%
%





\end{document}